\documentclass[%
reprint,
amsmath,amssymb,
aps,
]{revtex4-2}

\usepackage{lipsum}
\usepackage{bm}
\usepackage{amssymb}
\usepackage{amstext}
\usepackage{amsmath}
\usepackage{mathrsfs}
\usepackage{graphicx}
\usepackage{color}
\usepackage{amsthm}
\usepackage{floatrow}
\usepackage{array}
\renewcommand\arraystretch{2}
\usepackage{listings}
\usepackage{booktabs}
\usepackage{hyperref}% add hypertext capabilities
\begin{document}
	\preprint{APS/123-QED}
	\title{Thermodynamics of percolation in interacting systems}% Force line breaks with \\
	\thanks{Correspondence should be addressed to Yang Tian and Pei Sun.}%

	\author{Yizhou Xu}%
	\email{xuyz23@mails.tsinghua.edu.cn}
	\altaffiliation[]{Department of Mathematical Sciences, Tsinghua University, Beijing, 100084, China.}
	
	\author{Pei Sun}%
	\email{peisun@tsinghua.edu.cn}
	\altaffiliation[]{Department of Psychology \& Tsinghua Laboratory of Brain and Intelligence, Tsinghua University, Beijing, 100084, China.}

 \author{Yang Tian}
	\email{tyanyang04@gmail.com \& tiany20@mails.tsinghua.edu.cn}
	\altaffiliation[]{Department of Psychology \& Tsinghua Laboratory of Brain and Intelligence, Tsinghua University, Beijing, 100084, China.}
	
	%\affiliation{
	% Lunar Base
	%\affiliation{%
	
	%\collaboration{CLEO Collaboration}%%\noaffiliation
	
	
\begin{abstract}
Interacting systems can be studied as the networks where nodes are system units and edges denote correlated interactions. Although percolation on network is a unified way to model the emergence and propagation of correlated behaviours, it remains unknown how the dynamics characterized by percolation is related to the thermodynamics of phase transitions. It is non-trivial to formalize thermodynamics for most complex systems, not to mention calculating thermodynamic quantities and verifying scaling relations during percolation. In this work, we develop a formalism to quantify the thermodynamics of percolation in interacting systems, which is rooted in a discovery that percolation transition is a process for the system to lose the freedom degrees associated with ground state configurations. We derive asymptotic formulas to accurately calculate entropy and specific heat under our framework, which enables us to detect phase transitions and demonstrate the Rushbrooke equality (i.e., $\alpha+2\beta+\gamma=2$) in six representative complex systems (e.g., Bernoulli and bootstrap percolation, classical and quantum synchronization, non-linear oscillations with damping, and cellular morphogenesis). These results suggest the general applicability of our framework in analyzing diverse interacting systems and percolation processes.
\end{abstract}
\maketitle

Interacting systems formed by units with correlated behaviours are important research objects in physics and related fields \cite{wilson1973mathematical,amari1977dynamics,bressloff2010metastable,bressloff2010stochastic,binder1986spin,acebron2005kuramoto,de2014synchronization,scianna2012multiscale,chen2007parallel,chiang2016glass}. By its nature, an interacting system can be represented by the dynamics on a network $G\left(V,E\right)$, where nodes in $V$ are system units and edges in $E$ denote correlated interactions. The existence of an edge between units $i$ and $j$ is non-trivially determined by $R_{ij}$, the $\left(i,j\right)$-th entity in a specific correlation matrix $R$ that describes the relations among system units. Percolation on network \cite{li2021percolation,duminil2018sixty} may be a natural formalization that characterizes different interacting systems as networks in a unified manner, whose key idea is to represent the propagation of correlated behaviours via edge occupation (i.e., each edge is occupied only if the associated units synchronously behave). See Sec. \ref{ASec1-1} in SM for a summary of percolation on network.

It is long hoped to bridge between the dynamics modelled by percolation on network and the thermodynamic formalization of phase transitions (i.e, relating the observable quantities of percolation to their thermodynamic counterparts) \cite{hassan2017entropy,radicchi2009explosive,ziff2010scaling,bastas2011explosive,fortuin1972random,hassan2016universality}. However, previous progress is limited to the classic cases where percolation is defined on lattices and is less applicable to characterizing complex interacting systems (i.e., units in a real complex system may not necessarily be placed on a regular, uniform, and fixed structure). Meanwhile, the proposed thermodynamics concepts in previous works (e.g., see entropy and specific heat in Ref. \cite{hassan2017entropy}) critically rely on numerical computations and lack analytic or, at least, asymptotic expressions. To a large extent, percolation on network remains as a mathematical theory.

In the spirit of developing statistical mechanics of network science \cite{de2016spectral,park2004statistical}, here we suggest a possible paradigm to define thermodynamic concepts in an explicit manner, which is generally applicable to different kinds of percolation and interacting systems. These concepts enable us to measure the loss of freedom degrees during percolation, detect percolation transition, and verify important scaling relations in diverse complex systems where classical thermodynamic analysis is difficult or inapplicable. 

\begin{figure}[b!]
\includegraphics[width=\columnwidth]{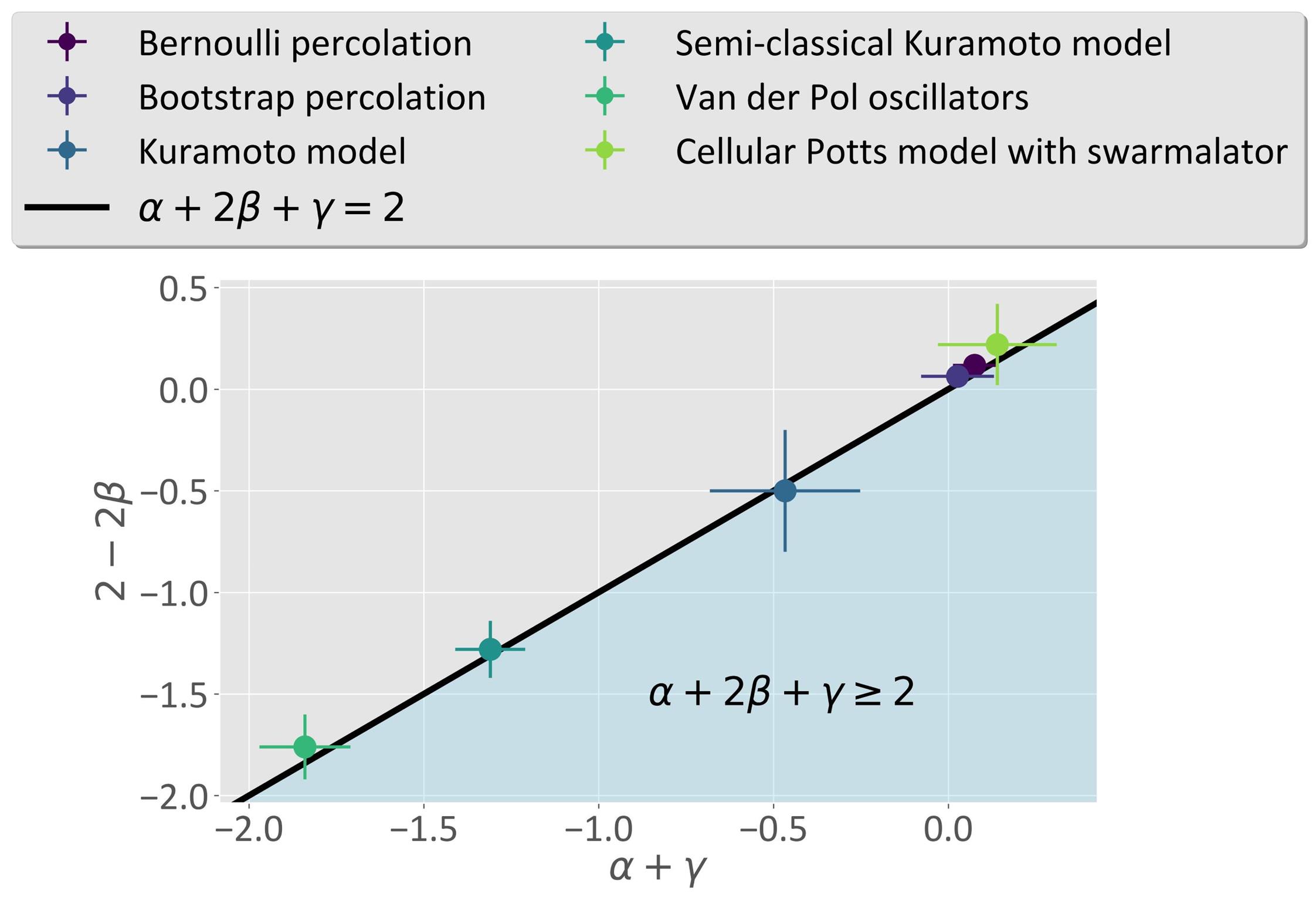}
\caption{\label{Fig1} Scaling relation verified in different systems, where error bars indicate the $95\%$ confidence interval of parameter estimation. The light blue area denotes the region where $\alpha+2\beta+\gamma\geq 2$ while the black line stands for $\alpha+2\beta+\gamma= 2$.} 
\end{figure}

Specifically, our theory is applied to study six representative complex systems: (1) the Bernoulli percolation on network \cite{georgakopoulos2018analyticity,li2021percolation,moore2000exact,newman2001random}; (2) the bootstrap percolation on network \cite{adler1988diffusion,baxter2010bootstrap}; (3) the classical synchronization described by the Kuramoto model \cite{rodrigues2016kuramoto,acebron2005kuramoto}; (4) the quantum synchronization emerged in a semi-classical Kuramoto model \cite{de2014synchronization}; (5) the non-linear oscillations with damping generated by a ring of Van der Pol oscillators \cite{van1928lxxii,peltola2007synthesis,ginoux2012van}; (6) the cellular aggregation and synchronization characterized by a mixture of the cellular Potts model \cite{unacellular,voss2018cellular} and the swarmalator \cite{sar2022dynamics,hong2023swarmalators}. Our framework enables us to analyze the scaling behaviours of order parameter $\text{OP}\propto \varepsilon^{\beta}$, specific heat $C\propto \varepsilon^{-\alpha}$, and susceptibility $\chi\propto \varepsilon^{-\gamma}$, when the difference between control parameter and critical point, $\varepsilon$, reduces to zero. We further verify the Rushbrooke inequality associated with these critical exponents, $\alpha+2\beta+\gamma\geq 2$ \cite{stanley1971phase}. Given the static scaling hypothesis, this inequality reduces to equality \cite{stanley1971phase}, which has been observed in diverse systems (e.g., see Ref. \cite{fytas2017restoration}). In our results, all considered complex systems obey the Rushbrooke inequality even though their model definitions are highly distinct. The actual values of critical exponents in all systems make $\alpha+2\beta+\gamma=2$ generally hold (see Fig. \ref{Fig1} for our main results), validating the applicability of our framework in thermodynamic analysis of percolation and interacting systems.

\paragraph*{Preliminaries.---} As the most general case, let us consider the occupation probability vector $\rho\in\left[0,1\right]^{E}$ determined by the function of correlation matrix $R$. Each component of $\rho$ denotes the occupation probability of a certain edge in a heterogeneous system. When the system is homogeneous, all components of $\rho$ are the same and $\rho$ reduces to a scalar. Given a $\rho$, we actually defines a probability space $\left(\{0,1\}^{ E},\sigma_{\eta} ,P_{\rho}\right)$ in which all percolation configurations live. Each percolation configuration $\mathbf{\eta}=\left(\eta_{e}:\;e=\left(i,j\right)\in E\right)\in\{0,1\}^{E}$ corresponds to a possible state of the propagation of correlated behaviours in the system, where $1$ denotes occupation (i.e., associated units exhibit correlated behaviours) and $0$ denotes non-occupation (i.e., associated units are currently uncorrelated). Please see Sec. \ref{ASec1-1} in SM for full definitions.

The essential difficulty lies in that statistical ensemble $\left(\{0,1\}^{E},\sigma_{\eta} ,P_{\rho}\right)$ can be highly non-trivial in most real interacting systems, where the function of correlation matrix $R$ is unknown and measuring the probability of each percolation configuration is impossible. In our work, we are no longer obsessed with the precise definition of $\rho$ and suggest an indirect way to study the properties of $\left(\{0,1\}^{E},\sigma_{\eta}, P_{\rho}\right)$ in the absence of \emph{a priori} knowledge. 

\paragraph*{The idea.---} Our idea is inspired by a simple fact. We consider a situation where there exist correlated behaviours modelled by a percolation process controlled by a latent $\rho$. We can always search through the system to find clusters even though the mechanism underlying percolation is unknown. For instance, let us assume that an arbitrary cluster of size $M$ is observed in the system (i.e., these $M$ units are connected following a certain wiring diagram). The actual wiring diagram of these $M$ units can be diverse because we have no constraint on $\rho$. However, these $M$ units must form a cluster (i.e., a connected component) as required by the percolation process. Therefore, the possibilities where these $M$ units are disconnected should be faint to avoid contradictions. 

As an opposite situation, we consider a neutral case without any propagation of correlated behaviours. There is no restraint on the wiring diagram of these $M$ units. In this case, all wiring diagrams are possible, including those where $M$ units do not form a cluster. 

Comparing between these two situations, one can immediately realize that the percolation process actually rejects some possibilities (i.e., freedom degrees) compared with the neutral state (i.e., all percolation configurations whose largest cluster sizes are smaller than $M$ have been excluded). The amount of lost freedom degrees during percolation lies in the heart of our theory. Because we primarily concern the lost quantity rather than detailed freedom degrees, our framework has less dependence on the knowledge about $\rho$ and becomes more applicable in complex systems.

\paragraph*{The theory.---} Below, we mathematically formalize the above idea. We first denote $\mathbf{H}\left(M\right)$ as the set of all possible percolation configurations (i.e., wiring diagrams) formed on the considered $M$ units (there are $2^{\binom {M} {2}}$ possibilities, see Sec. \ref{ASec1-2} in SM for instances). 

Then, we define $\left( \mathbf{H}\left(M\right),\sigma_{\eta},P_{\text{free}}\right)$ as the statistical ensemble corresponding to the neutral situation of these $M$ units. Every microstate $\eta$ (i.e., a percolation configuration) in this ensemble is possible to occur because the system is free of any constraint, leading us to the postulate of \emph{equal a priori probabilities}, i.e., $P_{\text{free}}\left(\eta\right)=2^{-\binom {M} {2}}$.

Meanwhile, we define $\left( \mathbf{H}\left(M\right),\sigma_{\eta} ,P_{\rho}\right)$, the probability space corresponding to the percolation process controlled by $\rho$. This is an ensemble where a percolation configuration is possible to occur if and only if the associated network of $M$ units is connected. Therefore, the probability of microstate $\eta$ is $P_{\rho}\left(\eta\right)= \frac{1}{Z_{\eta}\left(M\right)}\delta\left(L\left(\eta\right),M\right)$, where $L\left(\eta\right)\leq M$ is the largest cluster in the network associated with $\eta$, notion $\delta\left(\cdot,\cdot\right)$ denotes the Kronecker delta function, and $Z_{\eta}\left(M\right)$ is a partition function. See Sec. \ref{ASec1-2} in SM for instances.

The lost freedom degrees during percolation has a probability distribution $P_{\text{loss}}\left(\cdot\right)$. The information quantity (i.e., the Shannon entropy \cite{cover1999elements}) contained in this distribution quantifies the difference between $P_{\text{free}}\left(\cdot\right)$ and $P_{\rho}\left(\cdot\right)$. In our theory, we find that this information quantity can be calculate by 
 \begin{align}
S_{\text{loss}}\left(M\right)=\binom {M} {2}\log\left(2\right)-\mathbb{D}\left(P_{\text{loss}}\Vert P_{\text{free}}\right),\label{EQ1}
\end{align}
where $\mathbb{D}\left(P_{\text{loss}}\Vert P_{\text{free}}\right)$ denotes the relative entropy \cite{cover1999elements} between $P_{\text{loss}}\left(\cdot\right)$ and $P_{\text{free}}\left(\cdot\right)$. Please see Sec. \ref{ASec1-2} in SM for detailed derivations of Eq. (\ref{EQ1}). As shown in Eq. (\ref{EQ1}), the amount of lost freedom degrees measured by $S_{\text{loss}}\left(M\right)$ varies inversely with $\mathbb{D}\left(P_{\text{loss}}\Vert P_{\text{free}}\right)$. Therefore, we can use the relative entropy $\mathbb{D}\left(P_{\text{loss}}\Vert P_{\text{free}}\right)$ to study the freedom degrees remaining in $P_{\rho}\left(\cdot\right)$ indirectly. 

\paragraph*{Thermodynamic meaning.---} What is the thermodynamic meaning of the above formalization? Although $P_{\text{free}}\left(\cdot\right)$ and $P_{\text{loss}}\left(\cdot\right)$ are initially proposed for percolation process, we find that they coincide with two thermodynamic states (see Sec. \ref{ASec1-3} in SM for derivations). 

Let us assume an abstract system associated with $P_{\text{loss}}\left(\cdot\right)$, which is at equilibrium with a temperature $T_{\text{loss}}$. This system has two abstract energy levels, which respectively correspond to the microstates counted by $Z_{\eta}\left(M\right)$ (i.e., the percolation configurations $\eta$ that satisfy $L\left(\eta\right)=M$) and the microstates excluded from $Z_{\eta}\left(M\right)$ (i.e., the percolation configurations $\eta$ with $L\left(\eta\right)<M$). We notice that $P_{\text{loss}}\left(\cdot\right)$ coincides with a situation where all microstates with the first energy level are equally likely to occur while those with the second energy level are unlikely to occur. We define $U_{\eta}$ as the energy level of microstate $\eta$. Without loss of generality, we further set $U_{\eta}= r_{1}\in\mathbb{R}^{+}$ if $L\left(\eta\right)=M$ and $U_{\eta}=r_{2}\in\mathbb{R}^{+}$ if $L\left(\eta\right)<M$, such that $r_{1}>r_{2}$. If we consider a Boltzmann form of the distribution of microstates
\begin{align}
P_{\text{loss}}\left(\eta\right)\simeq\frac{1}{Z_{\text{loss}}}\exp\left(-\beta_{\text{loss}}U_{\eta}\right)\label{EQ2}
\end{align}
with an inverse temperature $\beta_{\text{loss}}=\frac{1}{T_{\text{loss}}}$ and a partition function $Z_{\text{loss}}$, this Boltzmann distribution should satisfy $\beta_{\text{loss}}\rightarrow \infty$ (i.e., the zero temperature limit) because a temperature tending to zero ensures 
\begin{align}
\lim_{\beta_{\text{loss}}\rightarrow \infty}P_{\text{loss}}\left(U_{\eta}=r_{1}\right)&=0,\label{EQ3}\\
\lim_{\beta_{\text{loss}}\rightarrow \infty}P_{\text{loss}}\left(U_{\eta}=r_{2}\right)&=\frac{1}{\vert \mathbf{H}\left(M\right)\vert-Z_{\eta}\left(M\right)},\label{EQ4}
\end{align}
to keep consistency with the property of $P_{\text{loss}}\left(\cdot\right)$. Therefore, we suggest to interpret the situation described by $P_{\text{loss}}\left(\cdot\right)$ as a zero temperature limit condition.

In a similar way, we can rewrite $P_{\text{free}}\left(\cdot\right)$ in a form of Boltzmann distribution  
\begin{align}
P_{\text{free}}\left(\eta\right)\simeq \frac{1}{Z_{\text{free}}}\exp\left(-\beta_{\text{free}}U_{\eta}\right),\label{EQ5}
\end{align}
which corresponds to the system at equilibrium with a temperature $T_{\text{free}}$. The property of $P_{\text{free}}\left(\cdot\right)$ requires each energy level $r\in\{r_{1},r_{2}\}$ to share the same probability, implying $\beta_{\text{free}}\rightarrow 0$ (i.e., the high temperature limit)
\begin{align}
\lim_{\beta_{\text{free}}\rightarrow 0}P_{\text{free}}\left(U_{\eta}=r\right)&=\frac{1}{\vert \mathbf{H}\left(M\right)\vert}.\label{EQ6}
\end{align}
Consequently, the situation described by $P_{\text{free}}\left(\cdot\right)$ can be treated as a high temperature limit condition.

Given the thermodynamic meanings of $P_{\text{loss}}\left(\cdot\right)$ and $P_{\text{free}}\left(\cdot\right)$, we turn to analyzing Eq. (\ref{EQ1}). Utilizing the properties of relative entropy and entropy \cite{cover1999elements}, we can reformulate $\mathbb{D}\left(P_{\text{loss}}\Vert P_{\text{free}}\right)$ as 
\begin{align}
\mathbb{D}\left(P_{\text{loss}}\Vert P_{\text{free}}\right)
=\Delta S-\frac{\Delta U}{T_{\text{free}}},\label{EQ7}
\end{align}
where $\Delta S=S\left(P_{\text{free}}\right)-S\left(P_{\text{loss}}\right)$ and $\Delta U=U_{\text{free}}-U_{\text{loss}}$. Notions $S\left(P_{\text{free}}\right)$ and $S\left(P_{\text{loss}}\right)$ denote the entropy values of the systems defined by Eq. (\ref{EQ2}) and Eq. (\ref{EQ5}), respectively. Meanwhile, notions $U_{\text{free}}$ and $U_{\text{loss}}$ measure the energies of these systems. Based on the high temperature limit condition in Eq. (\ref{EQ5}), we can derive
\begin{align}
\lim_{T_{\text{free}}\rightarrow\infty}\mathbb{D}\left(P_{\text{loss}}\Vert P_{\text{free}}\right)=\Delta S,\label{EQ8}
\end{align}
meaning that $\mathbb{D}\left(P_{\text{loss}}\Vert P_{\text{free}}\right)$ actually measures the entropy difference between the equilibrium states at zero and high temperature limits in our framework. This property supports us to apply $\mathbb{D}\left(P_{\text{loss}}\Vert P_{\text{free}}\right)$ to studying the thermodynamics underlying percolation.

Derivations of Eqs. (\ref{EQ2}-\ref{EQ8}) are shown in Sec. \ref{ASec1-3} in SM.

What does temperature mean for percolation? As percolation process approaches to its transition point, increasingly more freedom degrees contained in $P_{\text{loss}}\left(\cdot\right)$ are lost. These lost freedom degrees, living at the zero temperature limit, correspond to the situations where all units are located at their ground states (i.e., being least-energetic). In this manner, the percolation transition described by our theory is consistent with conventional thermodynamic phase transitions (e.g., the order-disorder transition in the Ising model) during which the probability mass of ground state configurations is reduced. As for $P_{\text{free}}\left(\cdot\right)$, a maximum entropy distribution, it corresponds to the most disorder behaviours of the system as temperature approaches to infinity.

\begin{figure*}[t!]
\includegraphics[width=\columnwidth]{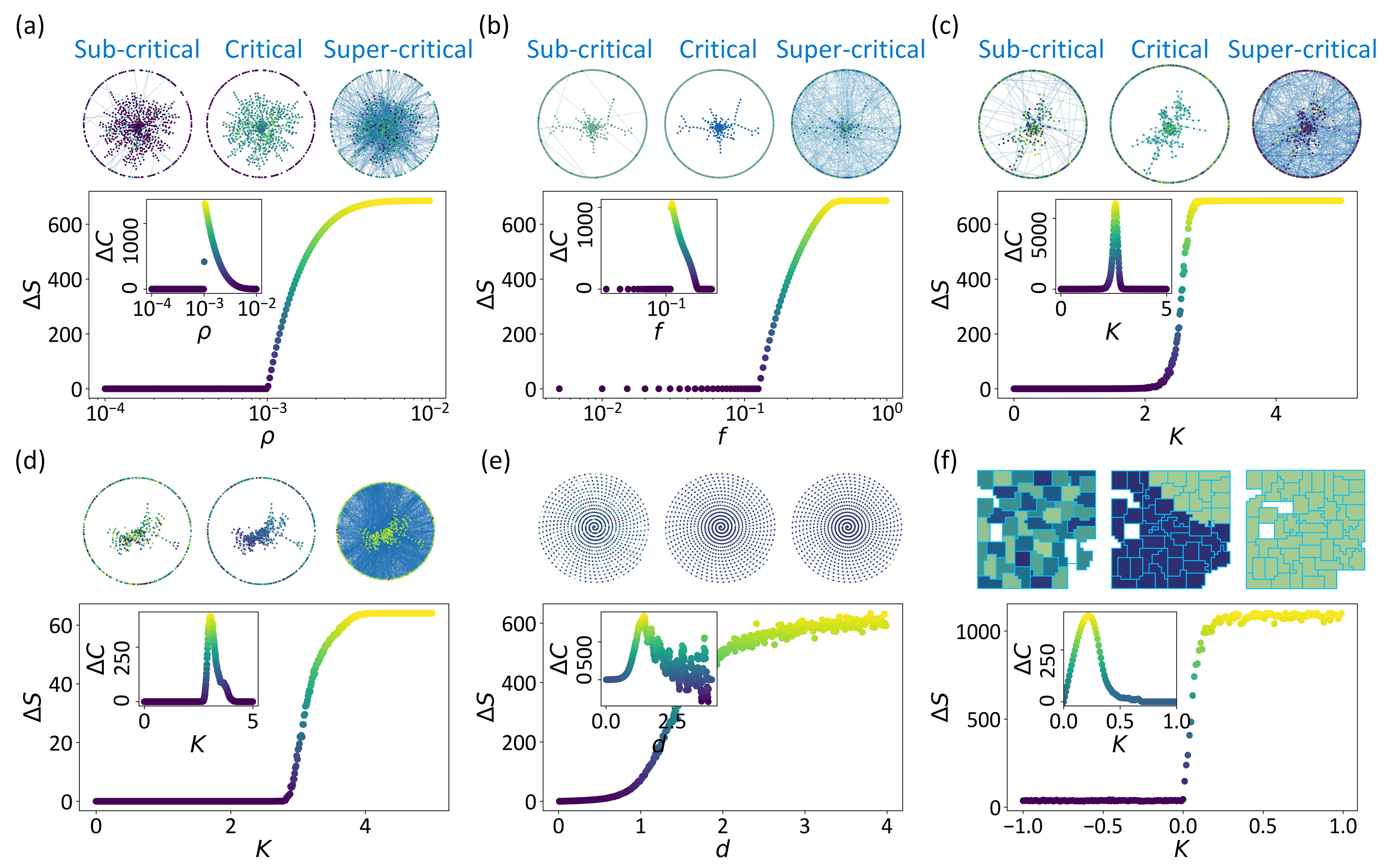}
\caption{\label{Fig2} The derived entropy, $\Delta S$, and specific heat, $\Delta C$, in the Bernoulli percolation (a), the bootstrap percolation (b), classical synchronization (c), quantum synchronization (d), non-linear oscillations with damping (e), and cellular morphogenesis (f). The instances of these systems are illustrated under sub-critical, critical, and super-critical conditions. Units are colored according to their degrees in (a, b, e) while they are colored according to phases in (c, d, f). Note that the ring of Van der Pol oscillators is shown with a spiral layout for visual clarity.} 
\end{figure*}

\paragraph*{The computation.---} The initial expression of relative entropy $\mathbb{D}\left(P_{\text{loss}}\Vert P_{\text{free}}\right)$, as shown in Sec. \ref{ASec1-2} in SM, relies on solving enumerative combinatorics problems that are NP hard. To avoid these limitations, we derive an asymptotic expression of $\mathbb{D}\left(P_{\text{loss}}\Vert P_{\text{free}}\right)$ using generating function tricks \cite{wilf2005generatingfunctionology,flajolet2009analytic}
\begin{align}
\mathbb{D}\left(P_{\text{loss}}\Vert P_{\text{free}}\right)=\left(M-1\right)\log\left(2\right)-\log\left(M\right)+o\left(1\right).\label{EQ9}
\end{align}
This asymptotic formula can accurately computes $\mathbb{D}\left(P_{\text{loss}}\Vert P_{\text{free}}\right)$ (see Sec. \ref{ASec1-4} in SM for accuracy demonstration) and is generally applicable to real systems with arbitrarily large sizes. At the thermodynamic limit, the entropy difference measured by $\mathbb{D}\left(P_{\text{loss}}\Vert P_{\text{free}}\right)$ asymptotically equals to an extensive property, $M\left(\log\left(2\right)+o\left(1\right)\right)$. For convenience, we set $M$ as the size of the giant cluster of correlated units in subsequent analysis.

\paragraph*{The experiment.---} To comprehensively validate our theory, we apply it to characterizing phase transitions, calculating critical exponents, and verifying scaling relations in interacting systems. We are especially interested in the cases where systems are highly complex (e.g., defined on irregular or variable structures) and classic thermodynamic analysis is less applicable (e.g., do not have solvable governing equations).

In our work, we have implemented the following systems: (1) the Bernoulli percolation on network \cite{georgakopoulos2018analyticity,li2021percolation,moore2000exact,newman2001random} whose control parameter is the occupation probability ($\rho$); (2) the bootstrap percolation on network \cite{adler1988diffusion,baxter2010bootstrap} controlled by the initial activation fraction ($f$); (3) the classical synchronization described by the Kuramoto model \cite{rodrigues2016kuramoto,acebron2005kuramoto}, whose control parameter is coupling strength ($K$); (4) the quantum synchronization emerged in a semi-classical Kuramoto model \cite{de2014synchronization}, whose control parameter is same as the classical case; (5) the non-linear oscillations with damping generated by Van der Pol oscillators \cite{van1928lxxii,peltola2007synthesis,ginoux2012van}, whose control parameter is interaction strength ($d$); (6) the cellular aggregation and synchronization characterized by a mixture of the cellular Potts model \cite{unacellular,voss2018cellular} and the swarmalator \cite{sar2022dynamics,hong2023swarmalators}, which shares control order parameter with classical synchronization. Among them, systems (1-4) exhibit dynamics in the high-dimensional structures represented by networks. System (5) are placed on a ring network and system (6) has variable structure during aggregation. Please see Sec. \ref{ASec2} in SM for model definitions.

For each system, we generate its $100$ realizations under every condition of control parameter to produce its phase transition behaviours (see Sec. \ref{ASec2} in SM for experiment settings). We generate networks for these realizations according to the propagation of correlation behaviours among units. The generated networks are used to measure $M$ and implement our theory (see Sec. \ref{ASec2} in SM for details). To precisely describe the phase transition as the distance between control parameter and transition point, $\varepsilon$, reduces to zero, we consider the following scaling behaviours
\begin{align}
     \Delta C \propto \varepsilon^{-\alpha},\;\;\;\;\;\;\;\;\;
     \text{OP} \propto \varepsilon^\beta,\;\;\;\;\;\;\;\;\;
     \chi\propto \varepsilon^{-\gamma},\label{EQ10}
\end{align}
where $\Delta C$ is the specific heat difference corresponding to $\Delta S$ in Eq. (\ref{EQ9}), order parameter (OP) denotes the giant cluster size during correlated behaviour propagation (here unit correlation is determined by model definition), and $\chi$ is the susceptibility (i.e., the first derivative of OP, which is equivalent to the mean size of finite clusters \cite{kirkpatrick1976percolation}). See Sec. \ref{ASec2} in SM for full definitions. After obtaining these critical exponents, we pursue to verify the Rushbrooke inequality 
\begin{align}
     \alpha+2\beta+\gamma\geq 2,\label{EQ11}
\end{align}
which is a well known scaling relation \cite{domb2000phase,stanley1971phase}. Under the static scaling hypothesis, we expect to see that the equality holds \cite{stanley1971phase}.

\paragraph*{Agreement between theory and experiment.---}A valid theory is expected to ensure the capacities of the derived $\Delta C$ and $\Delta S$ to characterize phase transitions. As shown in Fig. \ref{Fig2}, sharp increases of $\Delta S$ can be seen in all considered systems during continuous phase transitions, irrespective of how distinct model definitions are. Meanwhile, clear peaks of $\Delta C$ can be found at phase transition points (determined according to the behaviours of OP). These results qualitatively validate our theory in phase transition detection. Compared with classic approaches, the progress of our framework lies in its flexibility for different systems and the capacity to describe diverse phase transitions via a unified analysis pipeline.

As another requirement, a valid theory should derive critical exponents that obey scaling relations. In our experiment, the statistical fitting of critical exponents on all systems realizes high accuracy (see Sec. \ref{ASec2} in SM for fitting accuracy), suggesting that the scaling behaviours described in Eq. (\ref{EQ10}) generally hold. As shown in Fig. \ref{Fig1}, the equality $\alpha+2\beta+\gamma= 2$ under static scaling hypothesis principally holds on all kinds of systems because it is located within the $95\%$ confidence interval of parameter estimation when we statistically fit $\alpha$, $\beta$, and $\gamma$ in Eq. (\ref{EQ10}) using the data near phase transitions (see Sec. \ref{ASec2} in SM for concrete values of critical exponents). Note that confidence interval measurement presented here is necessary because the results of statistical regression, affected by random seeds of initialization, numerical errors, and the number of realizations, involve with  statistical uncertainties. To our best knowledge, this is the first work that verifies the Rushbrooke equality (or inequality) on the complex systems modelled by the bootstrap percolation on network \cite{adler1988diffusion,baxter2010bootstrap}, the Kuramoto model \cite{rodrigues2016kuramoto,acebron2005kuramoto}, the semi-classical Kuramoto model \cite{de2014synchronization}, the Van der Pol oscillator \cite{van1928lxxii,peltola2007synthesis,ginoux2012van}, and a cellular Potts model \cite{unacellular,voss2018cellular} defined with swarmalators \cite{sar2022dynamics,hong2023swarmalators}. Although the Bernoulli percolation on square and weighted planar stochastic lattices is reported to obey the Rushbrooke equality in Ref. \cite{hassan2017entropy}, our results serve as a generalization on networks. 

\paragraph*{Conclusion.---}In summary, we have presented a formalization to study the thermodynamics of percolation in interacting systems. Specifically, when the propagation of correlation behaviours among units is modelled by percolation on networks, our framework helps to analyze the loss of freedom degrees during percolation. Consistent with classic thermodynamic analysis, these lost freedom degrees are shown to be related to the probability mass of ground state configurations. Consequently, our analysis of lost freedom degrees offers an opportunity to define thermodynamic concepts during percolation. We have proposed asymptotic formulas to accurately calculate entropy and specific heat under our framework. As shown in our computational experiments on six complex systems, our theory provides a unified pipeline to detect and characterize phase transitions, where the derived critical exponents generally satisfy the Rushbrooke equality. Certainly, one is suggested to choose classic thermodynamic analysis when the target system is solvable and simple enough (e.g., with a completely known partition function). Our theory may be considered as an alternative when systems are highly intricate and classic analysis is inapplicable.

\paragraph*{Acknowledgements.---}This project is supported by the Artificial and General Intelligence Research Program of Guo Qiang Research Institute at Tsinghua University (2020GQG1017) as well as the Tsinghua University Initiative Scientific Research Program. 

 \bibliography{apssamp}
	\newpage
	% Produces the bibliography via BibTeX.
\appendix

\renewcommand{\figurename}{SFig.}
\setcounter{figure}{0}
\setcounter{equation}{0}
\section*{Supplementary materials}
\subsection{Thermodynamics underlying percolation}\label{ASec1}
\subsubsection{General definition of percolation on network}\label{ASec1-1}
The key idea of characterizing interacting systems by percolation on networks \cite{li2021percolation,duminil2018sixty} is to generate an all-to-all network of units (i.e., a complete graph) during initialization. Then each edge is occupied if and only if the associated units synchronously behave (i.e., these two units are correlated). All unoccupied edges are removed from the initial network, after which a filtered network is obtained to describe the propagation of correlated behaviours. In a probabilistic manner, we can consider an occupation probability vector
	\begin{align}
		\rho=h\left(R\right)=\left(\rho_{e}:\;e=\left(i,j\right)\in E\right)\in\left[0,1\right]^{E},\label{AEQ1}
	\end{align}
	where $h\left(\cdot\right)$ is a non-trivial element-wise function that maps correlation matrix $C$ to occupation probability vector $\rho$. A percolation configuration 
	\begin{align}
		\mathbf{\eta}=\left(\eta_{e}:\;e=\left(i,j\right)\in E\right)\in\{0,1\}^{E}\label{AEQ2}
	\end{align}
	lives in a probability space $\left(\{0,1\}^{ E},\sigma_{\eta} ,P_{\rho}\right)$, where $1$ denotes occupation (i.e., the edge is kept) and $0$ denotes non-occupation (i.e., the edge is deleted). Notion $\sigma_{\eta}$ denotes a $\sigma$-algebra and $P_{\rho}$ is a probability measure characterized by vector $\rho$. The propagation of correlated behaviours within the system is fully characterized by $\mathbb{P}_{\rho}$, based on which we can apply percolation theory to study the emergence of a giant cluster of correlated units or the size distribution of all finite clusters (note that the cluster is another name of the connected component) \cite{li2021percolation,georgakopoulos2018analyticity,duminil2018sixty}

 By adjusting the definition of $h\left(\cdot\right)$ for deriving different types of $\rho$, we can apply this framework to describing diverse kinds of correlated behaviour propagation, such as those characterized by Bernoulli percolation \cite{georgakopoulos2018analyticity}, explosive percolation \cite{matsoukas2015abrupt,boccaletti2016explosive}, and bootstrap percolation \cite{adler1988diffusion,baxter2010bootstrap}. For instance, we can define a simple $h\left(\cdot\right)$ to map $C$ to a scalar within $\left[0,1\right]$ that functions as a global occupation probability in Bernoulli percolation \cite{georgakopoulos2018analyticity}.

\subsubsection{A inforamtional view of freedom degree loss}\label{ASec1-2}
Let us consider a general case where the propagation of correlated behaviours in an interacting systems is modelled by an arbitrary percolation process. Given a $\rho$, we can search through the system to find emerged clusters. We assume that there exists an arbitrary cluster of size $M$ in the system (i.e., we require these $M$ units are connected following a certain wiring diagram). 

For these $M$ units, their wiring diagram can be diverse but must form a cluster (i.e., a connected component) as required by the percolation process. The possibilities where these $M$ units are disconnected should be faint. As an opposite case, we can consider a neutral situation where the propagation of correlated behaviours does not exist and there is no any restraint on the wiring diagram of these $M$ units. In this case, all wiring diagrams are possible, including those where $M$ units no longer form a cluster. Therefore, the percolation process actually rejects some possibilities (i.e., freedom degrees) compared with the neutral situation.

\begin{figure*}[t!]
\includegraphics[width=\columnwidth]{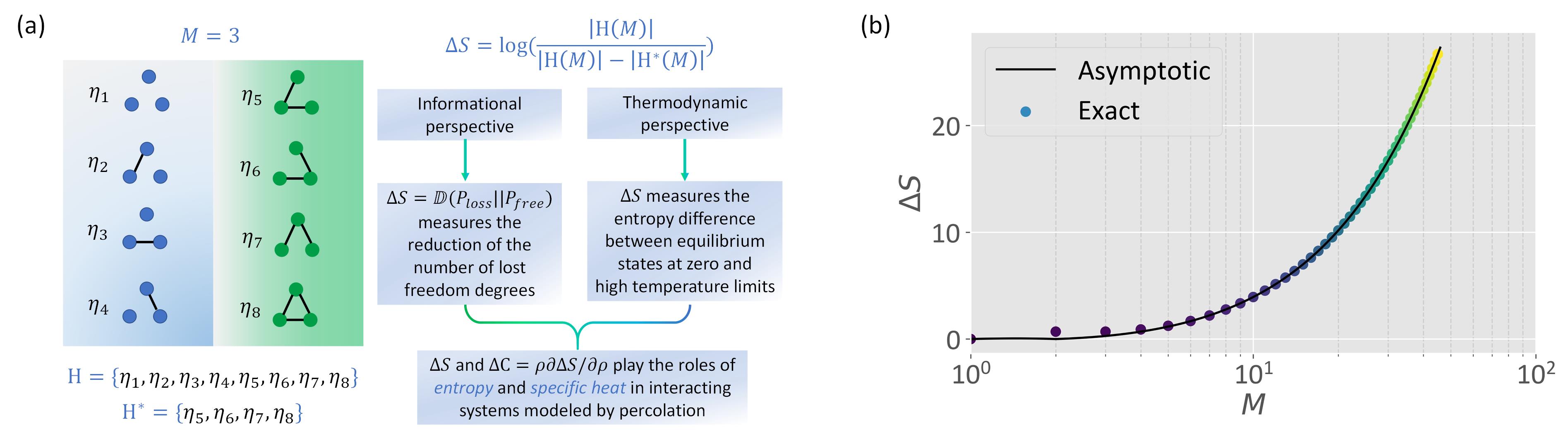}
\caption{\label{G1} Conceptual illustration of the informational and thermodynamic views of percolation in interacting systems. (a) presents an instance of the percolation configurations (i.e., microstates) contained in space $\mathbf{H}\left(3\right)$ and $\mathbf{H}^{*}\left(3\right)$, where we denote $\mathbf{H}^{*}\left(3\right)$ as the space of percolation configurations $\eta$ with $L\left(\eta\right)=3$ (left part). Meanwhile, a summary of our theoretical framework is provided (right part). (b) shows $\Delta S$ as a function of $M$ for illustration. The asymptotic values are obtained following Eq. (\ref{AEQ33}) while the exact values are obtained by numerically calculating Eq. (\ref{AEQ29}).} 
\end{figure*}

Below, we suggest a possible way to mathematically formalize the above analysis. The considered $M$ units can form $2^{\binom {M} {2}}$ kinds of wiring diagrams in total. Equivalently, there exist the same number of possible percolation configurations. We denote $ \mathbf{H}\left(M\right)$ as the set of these percolation configurations such that $\vert  \mathbf{H}\left(M\right)\vert=2^{\binom {M} {2}}$ (see SFig. \ref{G1}(a) for an instance).

We define $\left( \mathbf{H}\left(M\right),\sigma_{\eta},P_{\text{free}}\right)$ to characterize the statistical ensemble corresponding to the neutral situation of these $M$ units. In this ensemble, every percolation configuration is possible to occur because the system is free of any constraint, which is naturally related to the maximum entropy situation. This property inspires us to apply a postulate of \emph{equal a priori probabilities} and define the probability distribution of each microstate (i.e., percolation configuration) $\eta$ 
\begin{align}
P_{\text{free}}\left(\eta\right)= \frac{1}{\vert \mathbf{H}\left(M\right)\vert}=2^{-\binom {M} {2}}.\label{AEQ3}
\end{align}

Then we turn to $\left( \mathbf{H}\left(M\right),\sigma_{\eta} ,P_{\rho}\right)$, the probability space corresponding to the percolation process controlled by $\rho$. In this ensemble, a percolation configuration is possible to occur if and only if the associated network of $M$ units is connected. For convenience, we define $L\left(\eta\right)\leq M$ as the largest cluster in the network associated with $\eta$. The probability of microstate $\eta$ is 
\begin{align}
    P_{\rho}\left(\eta\right)= \frac{1}{Z_{\eta}\left(M\right)}\delta\left(L\left(\eta\right),M\right),\label{AEQ4}
\end{align}
where $\delta\left(\cdot,\cdot\right)$ denotes the Kronecker delta function and $Z_{\eta}\left(M\right)$ is a partition function
\begin{align}
    Z_{\eta}\left(M\right)=\sum_{\eta\in \mathbf{H}\left(M\right)}\delta\left(L\left(\eta\right),M\right) .\label{AEQ5}
\end{align}
See SFig. \ref{G1}(a) for the instances of $\eta$ with $L\left(\eta\right)= M$.

 Compared with $P_{\text{free}}\left(\cdot\right)$, lots of freedom degrees been reduced in $P_{\rho}\left(\cdot\right)$ because all percolation configurations whose largest cluster sizes are smaller than $M$ have been excluded. The freedom degree loss stands for the difference between these two distributions, to analyze which, we suggest to consider a re-normalized uniform probability distribution of excluded freedom degrees
 \begin{align}
P_{\text{loss}}\left(\eta\right)= \frac{1}{\sum_{\eta\in \mathbf{H}\left(M\right)}\mathbf{1}_{\left[0,M\right)}\left(L\left(\eta\right)\right)}\mathbf{1}_{\left[0,M\right)}\left(L\left(\eta\right)\right),\label{AEQ6}
\end{align}
where notion $\mathbf{1}_{A}\left(\cdot\right)$ denotes the indicator function defined on set $A$. To study the property of $P_{\text{loss}}\left(\cdot\right)$, we can use $P_{\text{free}}\left(\cdot\right)$, the maximum entropy distribution \cite{cover1999elements} over $ \mathbf{H}\left(M\right)$, as a reference to analyze how $P_{\text{loss}}\left(\cdot\right)$ approaches to or departs from the maximum entropy state. To realize this analysis, we derive the relative entropy between $P_{\text{loss}}\left(\cdot\right)$ and $P_{\text{free}}\left(\cdot\right)$ according to the property of relative entropy between uniform random variables \cite{cover1999elements}
\begin{align}
    \mathbb{D}\left(P_{\text{loss}}\Vert P_{\text{free}}\right)= \log\left(\frac{\vert \mathbf{H}\left(M\right)\vert}{\vert \mathbf{H}\left(M\right)\vert-Z_{\eta}\left(M\right)}\right).\label{AEQ7}
\end{align}

A direct benefit of analyzing $\mathbb{D}\left(P_{\text{loss}}\Vert P_{\text{free}}\right)$ lies in that it can be used to derive the Shannon entropy of distribution $P_{\text{loss}}\left(\cdot\right)$ \cite{cover1999elements}
 \begin{align}
S_{\text{loss}}\left(M\right)=\log\left(n\right)-\mathbb{D}\left(P_{\text{loss}}\Vert P_{U}\right),\label{AEQ8}
\end{align}
where $n$ is the total number of possibilities in the space of $P_{\text{loss}}\left(\cdot\right)$ and $P_{U}\left(\cdot\right)$ denotes the uniform distribution defined on these possibilities. In our situation, it is clear that all possibilities are included in set $ \mathbf{H}\left(M\right)$ and the uniform distribution defined on $ \mathbf{H}\left(M\right)$ coincides with $P_{\text{free}}\left(\cdot\right)$. Therefore, Eq. (\ref{AEQ8}) can be reformulated as
\begin{align}
S_{\text{loss}}\left(M\right)=\binom {M} {2}\log\left(2\right)-\mathbb{D}\left(P_{\text{loss}}\Vert P_{\text{free}}\right).\label{AEQ9}
\end{align}
According to Eq. (\ref{AEQ9}), as $\mathbb{D}\left(P_{\text{loss}}\Vert P_{\text{free}}\right)$ decreases (i.e., as $P_{\text{loss}}\left(\cdot\right)$ approaches to the maximum entropy distribution), the number of lost freedom degrees measured by $S_{\text{loss}}\left(M\right)$ increases. In an opposite case, as $\mathbb{D}\left(P_{\text{loss}}\Vert P_{\text{free}}\right)$ increases, the number of lost freedom degrees in $P_{\rho}\left(\cdot\right)$ compared with $P_{\text{free}}\left(\cdot\right)$ becomes small, leading to more freedom degrees remaining in $P_{\rho}\left(\cdot\right)$. In sum, although it is non-trivial to derive an analytic expression of $S_{\rho}\left(M\right)$, the number of freedom degrees remaining in $P_{\rho}\left(\cdot\right)$, we can use the relative entropy $\mathbb{D}\left(P_{\text{loss}}\Vert P_{\text{free}}\right)$ to study the evolution of $S_{\rho}\left(M\right)$ indirectly.

To use the presented framework to analyze percolation in practice, one needs to find an appropriate definition of $M$ depending on $\rho$. In our work, we suggest defining $M$ as the size of the giant cluster for convenience. This definition enables us to study the freedom degree loss related to the emergence of the giant cluster during percolation.

\subsubsection{A thermodynamics view of freedom degree loss}\label{ASec1-3}
Here we suggest one simple way to identify the role of $\mathbb{D}\left(P_{\text{loss}}\Vert P_{\text{free}}\right)$ in thermodynamics. 

We assume that the system associated with $\left( \mathbf{H}\left(M\right),\sigma_{\eta},P_{\text{loss}}\right)$ is at equilibrium with a temperature $T_{\text{loss}}$. This system has two abstract energy levels, which respectively correspond to the microstates counted by $Z_{\eta}\left(M\right)$ (i.e., the percolation configurations $\eta$ that satisfy $L\left(\eta\right)=M$) and the microstates excluded from $Z_{\eta}\left(M\right)$ (i.e., the percolation configurations $\eta$ with $L\left(\eta\right)<M$). Given this definition, we can see that Eq. (\ref{AEQ6}) coincides with a situation where all microstates with the first energy level are equally likely to occur while those with the second energy level are unlikely to occur. Inspired by this property, we can rewrite Eq. (\ref{AEQ6}) in a form of Boltzmann distribution 
\begin{align}
    P_{\text{loss}}\left(\eta\right)\simeq\frac{1}{Z_{\text{loss}}}\exp\left(-\beta_{\text{loss}}U_{\eta}\right),\label{AEQ10}
\end{align}
where $\beta_{\text{loss}}=\frac{1}{T_{\text{loss}}}$, notion $U_{\eta}$ denotes the energy level of microstate $\eta$, and $Z_{\text{loss}}=\sum_{\eta\in \mathbf{H}\left(M\right)}\exp\left(-\beta_{\text{loss}}U_{\eta}\right)$. Without loss of generality, we define $U_{\eta}= r_{1}\in\mathbb{R}^{+}$ if $L\left(\eta\right)=M$ and $U_{\eta}=r_{2}\in\mathbb{R}^{+}$ if $L\left(\eta\right)<M$, such that $r_{1}>r_{2}$. To keep consistency between Eq. (\ref{AEQ6}) and Eq. (\ref{AEQ10}) (i.e., make all microstates follow the lower energy level $r_{2}$ such that Eq. (\ref{AEQ6}) holds), we suggest to define $\beta_{\text{loss}}\rightarrow \infty$ (i.e., the zero temperature limit). In this case, we can have
\begin{align}
\lim_{\beta_{\text{loss}}\rightarrow \infty}P_{\text{loss}}\left(U_{\eta}=r_{1}\right)&=\lim_{\beta_{\text{loss}}\rightarrow \infty}\frac{\exp\left(-\beta_{\text{loss}}r_{1}\right)}{Z_{\text{loss}}},\label{AEQ11}\\&=0,\label{AEQ12}
\end{align}
and
\begin{align}
\lim_{\beta_{\text{loss}}\rightarrow \infty}P_{\text{loss}}\left(U_{\eta}=r_{2}\right)&=\lim_{\beta_{\text{loss}}\rightarrow \infty}\frac{\exp\left(-\beta_{\text{loss}}r_{2}\right)}{Z_{\text{loss}}},\label{AEQ13}\\&=\frac{1}{\vert \mathbf{H}\left(M\right)\vert-Z_{\eta}\left(M\right)},\label{AEQ14}
\end{align}
which are consistent with the distribution in Eq. (\ref{AEQ6}). Therefore, we can interpret the situation described by Eq. (\ref{AEQ6}) and Eq. (\ref{AEQ10}) as a zero temperature limit condition.

Similarly, we can rewrite Eq. (\ref{AEQ3}) in a form of Boltzmann distribution  
\begin{align}
P_{\text{free}}\left(\eta\right)\simeq \frac{1}{Z_{\text{free}}}\exp\left(-\beta_{\text{free}}U_{\eta}\right),\label{AEQ15}
\end{align}
which corresponds to the system at equilibrium with a temperature $T_{\text{free}}$. In Eq. (\ref{AEQ15}), we denote $Z_{\text{free}}=\sum_{\eta\in \mathbf{H}\left(M\right)}\exp\left(-\beta_{\text{free}}U_{\eta}\right)$. To ensure that Eq. (\ref{AEQ15}) is consistent with Eq. (\ref{AEQ3}), we suggest to define $\beta_{\text{free}}\rightarrow 0$ (i.e., the high temperature limit). This definition enables each energy level $r\in\{r_{1},r_{2}\}$ to share the same probability
\begin{align}
\lim_{\beta_{\text{free}}\rightarrow 0}P_{\text{free}}\left(U_{\eta}=r\right)&=\lim_{\beta_{\text{free}}\rightarrow 0}\frac{\exp\left(-\beta_{\text{free}}r\right)}{Z_{\text{free}}},\label{AEQ16}\\&=\frac{1}{\vert \mathbf{H}\left(M\right)\vert},\label{AEQ17}
\end{align}
which is consistent with Eq. (\ref{AEQ3}). Therefore, we suggest that the situation described by Eq. (\ref{AEQ3}) and Eq. (\ref{AEQ15}) is a high temperature limit condition.

According to the property of relative entropy \cite{cover1999elements}, we can express $\mathbb{D}\left(P_{\text{loss}}\Vert P_{\text{free}}\right)$ as
\begin{align}
    \mathbb{D}\left(P_{\text{loss}}\Vert P_{\text{free}}\right)=S\left(P_{\text{loss}},P_{\text{free}}\right)-S\left(P_{\text{loss}}\right),\label{AEQ18}
\end{align}
where $S\left(P_{\text{loss}},P_{\text{free}}\right)$ denotes the cross entropy 
\begin{align}
    &S\left(P_{\text{loss}},P_{\text{free}}\right)\notag\\=&-\sum_{\eta\in \mathbf{H}\left(M\right)}P_{\text{loss}}\left(\eta\right)\log\left(P_{\text{free}}\left(\eta\right)\right),\label{AEQ19}\\
    =&\log\left(Z_{\text{free}}\right)+\frac{\beta_{\text{free}}}{Z_{\text{loss}}}\left(\sum_{\eta\in \mathbf{H}\left(M\right)}\exp\left(-\beta_{\text{loss}}U_{\eta}\right)U_{\eta}\right),\label{AEQ20}\\
    =&\log\left(Z_{\text{free}}\right)+\beta_{\text{free}}U_{\text{loss}}.\label{AEQ21}
\end{align}
In Eq. (\ref{AEQ21}), term $U_{\text{loss}}$ denotes the energy of the system described by Eq. (\ref{AEQ10}). Because we can derive entropy $S\left(P_{\text{free}}\right)$ as
\begin{align}
S\left(P_{\text{free}}\right)=\log\left(Z_{\text{free}}\right)+\beta_{\text{free}}U_{\text{free}},\label{AEQ22}
\end{align}
we can combine Eqs. (\ref{AEQ21}-\ref{AEQ22}) to further obtain 
\begin{align}
S\left(P_{\text{loss}},P_{\text{free}}\right)=&S\left(P_{\text{free}}\right)-\beta_{\text{free}}\left(U_{\text{free}}-U_{\text{loss}}\right).\label{AEQ23}
\end{align}
Finally, we can rewrite Eq. (\ref{AEQ18}) as
\begin{align}
&\mathbb{D}\left(P_{\text{loss}}\Vert P_{\text{free}}\right)\notag\\=&S\left(P_{\text{free}}\right)-S\left(P_{\text{loss}}\right)-\beta_{\text{free}}\left(U_{\text{free}}-U_{\text{loss}}\right),\label{AEQ24}\\
=&\Delta S-\frac{\Delta U}{T_{\text{free}}},\label{AEQ25}
\end{align}
where $\Delta S=S\left(P_{\text{free}}\right)-S\left(P_{\text{loss}}\right)$ and $\Delta U=U_{\text{free}}-U_{\text{loss}}$ are used in Eq. (\ref{AEQ25}) (please note that $\Delta$ stands for numerical difference rather than derivative). Eq. (\ref{AEQ25}) may suggest the role of $\mathbb{D}\left(P_{\text{loss}}\Vert P_{\text{free}}\right)$ in thermodynamics. Given the high temperature limit condition in Eq. (\ref{AEQ16}), we can readily derive
\begin{align}
\lim_{T_{\text{free}}\rightarrow\infty}\mathbb{D}\left(P_{\text{loss}}\Vert P_{\text{free}}\right)=\Delta S,\label{AEQ26}
\end{align}
meaning that the concerned relative entropy $\mathbb{D}\left(P_{\text{loss}}\Vert P_{\text{free}}\right)$ actually measures the entropy difference between the equilibrium states at low and high temperature limits in our case. This property supports us to apply $\mathbb{D}\left(P_{\text{loss}}\Vert P_{\text{free}}\right)$ to studying the thermodynamics underlying percolation.

\subsubsection{Asymptotic expression of entropy difference}\label{ASec1-4}

To support a practical analysis of the entropy difference discussed in Sec. \ref{ASec1-3}, we present an asymptotic expression of $\mathbb{D}\left(P_{\text{loss}}\Vert P_{\text{free}}\right)$ based on several theorems in graph theory.

Let us consider $\{2^{\binom {M} {2}}\}$ and $\{Z_{\eta}\left(M\right)\}$ with $M\in\mathbb{N}^{+}$, which are the series of the total number of percolation configurations (i.e., microstates) in $ \mathbf{H}\left(M\right)$ and the series of the number of percolation configurations whose largest cluster sizes are $M$, respectively. From a graph theory perspective, we are equivalently analyzing the series of the number of graphs and the series of the number of connected graphs formed on $M$ labelled units. As suggested by Refs. \cite{wilf2005generatingfunctionology,flajolet2009analytic}, an exponential formula (also referred to as the polymer expansion in physics) can bridge between the exponential generating functions of these two series
\begin{align}
\sum_{M=1}^{\infty}Z_{\eta}\left(M\right)\frac{x^{M}}{M!}=\log\left(1+\sum_{M=1}^{\infty}2^{\binom {M} {2}}\frac{x^{M}}{M!}\right),\label{AEQ27}
\end{align}
We can expand the right side of Eq. (\ref{AEQ27}) via the Taylor expansion 
\begin{align}
\sum_{M=1}^{\infty}Z_{\eta}\left(M\right)\frac{x^{M}}{M!}=\sum_{n=0}^{\infty}\frac{\left(-1\right)^{n}}{n+1}\left(\sum_{M=1}^{\infty}2^{\binom {M} {2}}\frac{x^{M}}{M!}\right)^{n+1}.\label{AEQ28}
\end{align}
Certainly, it is impossible to analytically solve Eq. (\ref{AEQ28}) to derive $Z_{\eta}\left(M\right)$. However, we can reorganize Eq. (\ref{AEQ28}) and compare between its left and right sides to find an appropriate expression of coefficient term $Z_{\eta}\left(M\right)$. Specifically, we can observe an intricate expression using the method proposed in Ref. \cite{flajolet2009analytic}
\begin{widetext}
\begin{align}
Z_{\eta}\left(M\right)&=2^{\binom {M} {2}}+\sum_{k=2}^{\infty}\frac{\left(-1\right)^{k+1}}{k}\left[\sum_{M_{1}+\ldots+M_{k}=M}\binom {M} {M_{1},\ldots,M_{k}}2^{\sum_{i=1}^{k}\binom {M_{i}} {2}}\right],\label{AEQ29}\\
&=2^{\binom {M} {2}}-\frac{1}{2}\left[\sum_{M_{1}+M_{2}=M}\binom {M} {M_{1},M_{2}}2^{\binom {M_{1}} {2}+\binom {M_{2}} {2}}\right]+o\left(-\frac{1}{2}\left[\sum_{M_{1}+M_{2}=M}\binom {M} {M_{1},M_{2}}2^{\binom {M_{1}} {2}+\binom {M_{2}} {2}}\right]\right),\label{AEQ30}
\end{align}
\end{widetext}
where $o\left(\cdot\right)$ denotes an infinitesimal of higher order. Here Eq. (\ref{AEQ30}) is derived from the fact that the first term in the summation in Eq. (\ref{AEQ29}) exhibits predominance in determining the result of summation. Moreover, after analyzing all possible combinations of $\left(M_{1},M_{2}\right)$ in the predominant term of Eq. (\ref{AEQ30}), we find that $\left(M_{1}=M-1,M_{2}=1\right)$ and $\left(M_{1}=1,M_{2}=M-1\right)$ are predominant while other combinations are more negligible. Therefore, we can approximate Eq. (\ref{AEQ30}) as
\begin{align}
Z_{\eta}\left(M\right)&=2^{\binom {M} {2}}\left(1-M2^{-M+1}+o\left(2^{-M+1}\right)\right).\label{AEQ31}
\end{align}

Based on Eq. (\ref{AEQ31}), we can derive an asymptotic expression of $\mathbb{D}\left(P_{\text{loss}}\Vert P_{\text{free}}\right)$
\begin{align}
\mathbb{D}\left(P_{\text{loss}}\Vert P_{\text{free}}\right)&=\log\left(\frac{1}{M2^{-M+1}-o\left(2^{-M+1}\right)}\right),\label{AEQ32}
\\
&=\left(M-1\right)\log\left(2\right)-\log\left(M\right)+o\left(1\right),\label{AEQ33}
\end{align}
which can be practically used in application. According to Eq. (\ref{AEQ33}), the entropy difference measured by $\mathbb{D}\left(P_{\text{loss}}\Vert P_{\text{free}}\right)$ asymptotically equals to $M\left(\log\left(2\right)+o\left(1\right)\right)$, serving as an extensive property at the thermodynamic limit. Please see SFig. \ref{G1}(b) for an illustration.

\begin{figure*}[t!]
\includegraphics[width=\columnwidth]{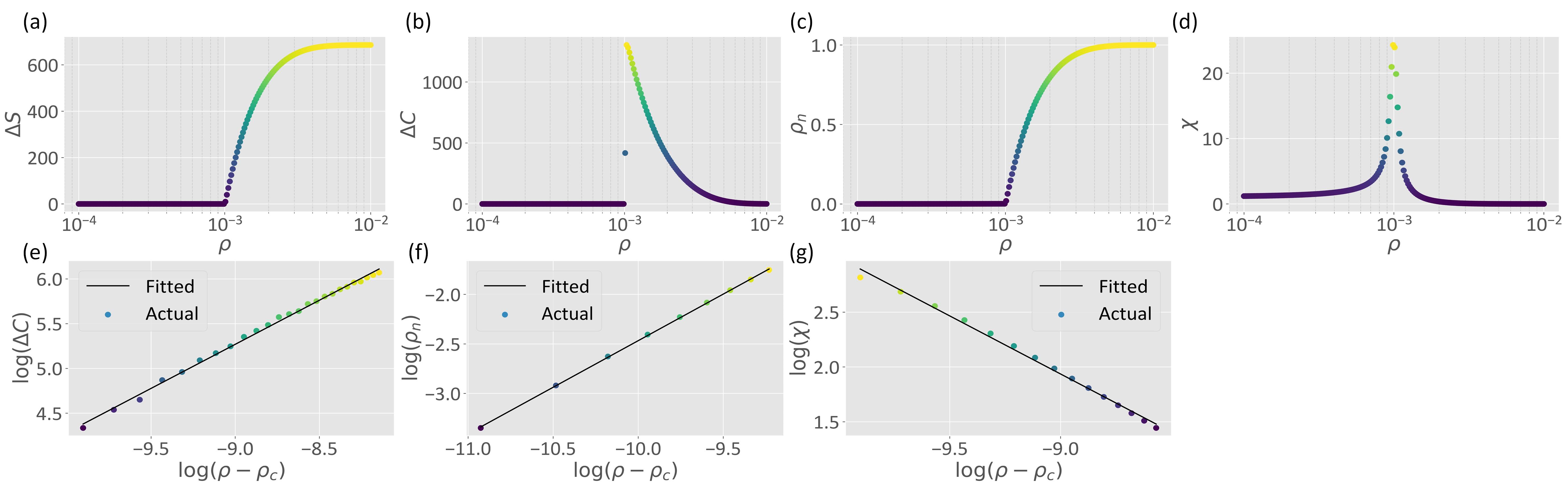}
\caption{\label{G2} Continuous phase transition modelled by the Bernoulli percolation. The derived $\Delta S$, $\Delta C$, $\rho_{n}$, and $\chi$ are presented as the functions of corresponding control parameters. Meanwhile, a linear fitting with least square method is applied to estimate the critical exponents defined in Eqs. (\ref{AEQ35}-\ref{AEQ37}) based on the data near critical points.} 
\end{figure*}

\begin{table*}[!t]
\centering
\renewcommand\arraystretch{1}
\small{
\begin{tabular}{lcccl}
\hline
 Phase transition model & $\alpha$                      & $\beta$                      & $\gamma$     & $\alpha+2\beta+\gamma$                                                       \\\hline
Bernoulli percolation & $-0.985\pm 0.022$ ($R^{2}=0.997$) & $0.941\pm 0.010$ ($R^{2}=0.9998$) & $1.06\pm 0.04$ ($R^{2}=0.995$) & $1.957\pm 0.082$  \\
 \hline
\end{tabular}
}
\caption{\label{Tab1}Critical exponents of the Bernoulli percolation. We report the estimated value $\pm$ the radius of $95\%$ confidence interval for each critical exponent, accompanied by the fitting accuracy measured by $R^{2}$.}
\end{table*}

\subsubsection{Summary of thermodynamics associated with percolation}\label{ASec1-5}
For a brief summary, we have suggested to consider $\mathbb{D}\left(P_{\text{loss}}\Vert P_{\text{free}}\right)$ as an measure of freedom degree loss during percolation in Sec. \ref{ASec1-2}, which indirectly measures the remaining freedom degrees in distribution $P_{\rho}\left(\cdot\right)$. From an information theory perspective, an increasing $\mathbb{D}\left(P_{\text{loss}}\Vert P_{\text{free}}\right)$ indicates the departure of $P_{\text{loss}}\left(\cdot\right)$ from the maximum entropy state, implying more freedom degrees in $P_{\rho}\left(\cdot\right)$. From a thermodynamics perspective shown in Sec. \ref{ASec1-3}, we suggest to understand $\mathbb{D}\left(P_{\text{loss}}\Vert P_{\text{free}}\right)$ as the entropy difference $\Delta S$ between the systems at low and
high temperature limits. In Sec. \ref{ASec1-4}, an asymptotic expression of this entropy difference is presented. SFig. \ref{G1} summarizes our key idea.

Based on the framework presented above, we mainly study the entropy difference itself and its corresponding specific heat difference
\begin{align}
\Delta C=\rho\frac{\partial }{\partial \rho}\left(M\left(\rho\right)-1\right)\log\left(2\right)-\log\left(M\left(\rho\right)\right)+o\left(1\right),\label{AEQ34}
\end{align}
where we explicitly indicate $M$ as a function of $\rho$ during percolation. For convenience, we primarily use $M$ as the size of the giant cluster of correlated units in our subsequent analysis. Similar to $\Delta S$, the proposed $\Delta C$ measures the difference of specific heat between the systems at zero and high temperature limits. As $\Delta C$ increases, the specific heat corresponding to distribution $P_{\rho}\left(\cdot\right)$ (i.e., remaining freedom degrees) is expected to decrease.

To precisely characterize the phase transition emerged during the correlated behaviour propagation modelled by percolation, we consider the following critical exponents  
\begin{align}
     \Delta C &\propto \vert\rho-\rho_{c}\vert^{-\alpha},\label{AEQ35}\\
     \rho_n&\propto \vert\rho-\rho_{c}\vert^\beta,\label{AEQ36}\\
     \chi&\propto \vert\rho-\rho_{c}\vert^{-\gamma},\label{AEQ37}
\end{align}
if the phase transition is continuous, where $\rho_{c}$ denotes the critical point, notion $\rho_{n}$ measures the probability for
a random node to belong to the giant cluster, and $\chi$ is the mean size of all finite clusters \cite{kirkpatrick1976percolation}. While Eqs. (\ref{AEQ36}-\ref{AEQ37}) have been extensively explored in percolation theories \cite{li2021percolation}, Eq. (\ref{AEQ35}) directly depends on our theory. After deriving these critical exponents, we pursue to verify the Rushbrooke inequality 
\begin{align}
     \alpha+2\beta+\gamma\geq 2,\label{AEQ38}
\end{align}
which is a well known scaling relation \cite{domb2000phase,stanley1971phase}. 

\subsection{Application on percolation and interacting systems}\label{ASec2}

\subsubsection{Bernoulli percolation process}\label{ASec2-1}

As the most simple instance, we first illustrate a Bernoulli percolation process on network controlled by occupation probability $\rho$ \cite{georgakopoulos2018analyticity}. The initial network is a complete graph with $\vert V\vert=1000$ units.
	
To measure $\Delta S$ and $\Delta C$, we need to derive $M$, the giant cluster size, as a function of $\rho$. According to the generating function approach \cite{li2021percolation}, the probability for a random node to belong to the giant cluster, $\rho_{n}$, and the probability for a random edge to belong to the giant cluster, $\rho_{e}$, obey the following relations
\begin{align}
		\rho_{e}&=1-\sum_{k}P_{k}\frac{k}{\sum_{k}kP_{k}}\left(1-\rho_{e}\rho\right)^{k},\label{AEQ39}\\
		\rho_{n}&=1-\sum_{k}P_{k}\left(1-\rho_{e}\rho\right)^{k},\label{AEQ40}
\end{align}
where $P_{k}$ denotes the probability of finding a node with $k$ degrees in the corresponding graph of the interacting system. In practice, one can first solve Eq. (\ref{AEQ39}) and insert $\rho_{e}$ into Eq. (\ref{AEQ40}) to derive $\rho_{n}$. Given a $\rho_{n}$, the size of giant cluster can be directly obtained as
\begin{align}
M=\lfloor\rho_{n}\vert V\vert\rfloor. \label{AEQ41}
\end{align} 
 These derivations enable us to calculate $\Delta S$ and $\Delta C$.

In SFigs. \ref{G2}(a-d), we show $\Delta S$, $\Delta C$, $\rho_{n}$ (the order parameter of the Bernoulli percolation), and $\chi$ (the mean size of finite clusters is numerically counted) as the functions of $\rho$ (the control parameter), respectively. It can be seen that our proposed $\Delta S$ and $\Delta C$ precisely capture the phase transition at the critical point (i.e., percolation transition threshold) $\rho_{c}$ \cite{li2021percolation,moore2000exact,newman2001random} 
	\begin{align}
		\rho_{c}&=\left(\frac{\partial }{\partial x} \sum_{k}P_{k}\frac{k}{\sum_{k}kP_{k}}x^{k-1}\Bigg\vert_{x=1}\right)^{-1},\label{AEQ42}\\
  &=\frac{1}{\vert V\vert-2}.\label{AEQ43}
	\end{align}
As the phase transition happens, the sharp increase of entropy difference $\Delta S$ suggests the rapid departure of distribution $P_{\text{loss}}\left(\cdot\right)$ from the maximum entropy state, implying that lost freedom degrees become increasingly negligible and $P_{\rho}\left(\cdot\right)$ contains most part of information of the statistical ensemble. The proposed specific heat difference $\Delta C$ diverges at the critical point, which meets the common behaviour of specific heat in most critical phenomena. Applying a log-log linear fitting on $\Delta C$, $\rho_{n}$, and $\chi$ near the critical point, we derive the concerned critical exponents, which are shown in Table. \ref{Tab1}. According to the estimated results, the Rushbrooke
inequality in Eq. (\ref{AEQ35}) generally holds because $2$ is covered by the $95\%$ confidence interval of the estimated $\alpha+2\beta+\gamma$.

\begin{figure*}[t!]
\includegraphics[width=\columnwidth]{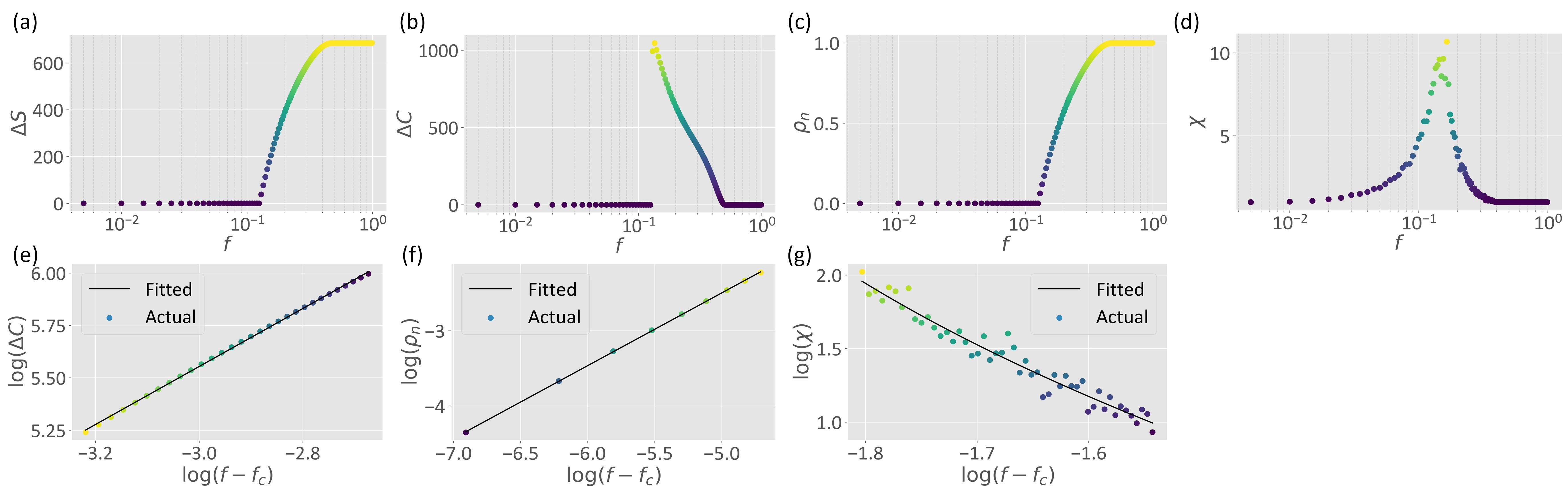}
\caption{\label{G3} Continuous phase transition modelled by the bootstrap percolation. The derived $\Delta S$, $\Delta C$, $\rho_{n}$, and $\chi$ are presented as the functions of corresponding control parameters. Meanwhile, a linear fitting with least square method is applied to estimate the critical exponents defined in Eqs. (\ref{AEQ35}-\ref{AEQ37}) based on the data near critical points. } 
\end{figure*}

\begin{table*}[!t]
\centering
\renewcommand\arraystretch{1}
\small{
\begin{tabular}{lcccl}
\hline
 Phase transition model & $\alpha$                      & $\beta$                      & $\gamma$     & $\alpha+2\beta+\gamma$                                                       \\\hline
Bootstrap percolation &  $-1.384\pm 0.011$ ($R^{2}=0.9995$) & $0.968\pm 0.007$ ($R^{2}=0.9999$) & $1.410\pm 0.093$  ($R^{2}=0.9999$) & $1.962\pm 0.118$  \\
 \hline
\end{tabular}
}
\caption{\label{Tab2}Critical exponents of the bootstrap percolation. We report the estimated value $\pm$ the radius of $95\%$ confidence interval for each critical exponent, accompanied by the fitting accuracy measured by $R^{2}$.}
\end{table*}

\subsubsection{Bootstrap percolation process}\label{ASec2-2}

We also consider the bootstrap percolation \cite{adler1988diffusion,baxter2010bootstrap}, a variant of the $k$-core percolation\cite{dorogovtsev2006k} that describes an activation processes, as a more complicated illustration of continuous phase transition. Specifically, we implement a bootstrap percolation controlled by an initial activation fraction $f$ in a random regular network of $1000$ units with degree $k$, where one node can be activated if at least $k^{*}$ of its neighbors are activated. 

For convenience, we consider the case where $k=3$ and $k^*=2$. Following Ref. \cite{di2019insights}, we need to first solve the following equations
\begin{align}
    \lambda&=\varepsilon-f\left(1-\lambda-\sigma\right)^{2}-\left(1-f\right)\left(\varepsilon-\lambda\right)^{2}\label{AEQ44},\\
\sigma&=2\left(1-f\right)\left[\varepsilon\left(1-\varepsilon\right)-\left(\varepsilon-\lambda\right)\left(1-\varepsilon-\sigma\right)\right],\label{AEQ45}
\end{align}
where $\varepsilon=\min\left(\frac{f}{1-f},1\right)$. Then, the order parameter (i.e.,  the probability for
a random node to belong to the giant cluster) is given as
\begin{align}
    \rho_n=&f+\left(1-f\right)\left[\varepsilon^3+3\varepsilon^2(1-\varepsilon+3\varepsilon\left(1-\varepsilon\right)^2\right]\notag\\&-f\left(1-\lambda-\sigma\right)^3\notag\\&-\left(1-f\right)\Big[\left(\varepsilon-\lambda\right)^3+3\left(1-\varepsilon-\sigma\right)\left(\varepsilon-\lambda\right)^2\notag\\&+3\left(\varepsilon-\lambda\right)\left(1-\varepsilon-\sigma\right)^2\Big].\label{AEQ46}
\end{align}

Based on these derivations, we can subsequently measure $M$ to calculate $\Delta S$, $\Delta C$, $\rho_{n}$, and $\chi$ (numerically counted as the mean size of finite clusters) as the functions of $f$. These results are presented in SFigs. \ref{G3}(a-d) in a way similar to our analysis of the Bernoulli percolation. As $f$ increases from $0$ to $1$, a phase transition appears around $f=0.125$, which can be derived by the marginal stability analysis at the fixed point $\left(\lambda,\sigma\right)=\left(0,0\right)$ of Eqs. (\ref{AEQ44}-\ref{AEQ45}). As shown in our results, the phase transition can be accurately captured by the trends of $\Delta S$ and $\Delta C$. By implementing the log-log linear fitting on the data near the critical point, we obtain the critical exponents shown in Table. \ref{Tab2} that satisfy the Rushbrooke
inequality in Eq. (\ref{AEQ35}).

\subsubsection{Classical synchronization}\label{ASec2-3}
Apart from the standard continuous phase transitions modelled by the Bernoulli percolation and the bootstrap percolation, we also consider more complicated propagation phenomena of correlated behaviours in complex interacting systems. In this section, we primarily focus on synchronization, a widespread phenomenon in diverse interacting systems, such as brains, social networks, and bird flocks \cite{boccaletti2002synchronization,arenas2008synchronization,boccaletti2016explosive,ghosh2022synchronized}.

\begin{figure*}[t!]
\includegraphics[width=\columnwidth]{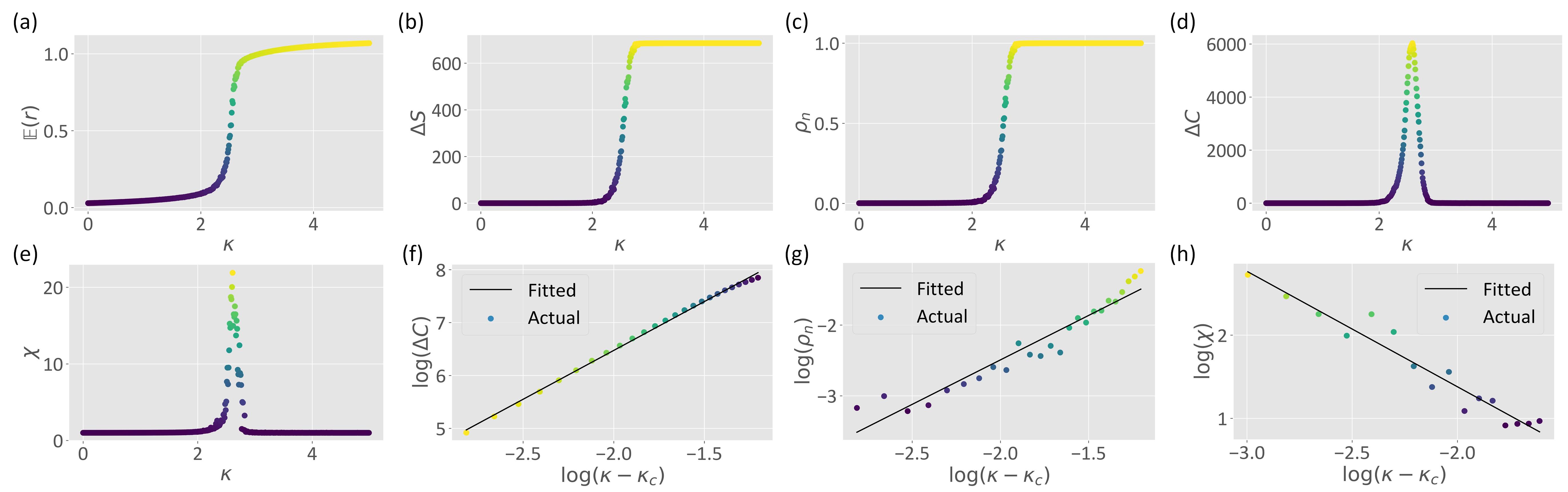}
\caption{\label{G4} Phase transition in the Kuramoto model. (a-e) The derived $\mathbb{E}\left(r\right)$, $\Delta S$, $\Delta C$, $\rho_{n}$, and $\chi$ are presented as the functions of $K$. (f-h) A linear fitting is applied to estimate the critical exponents in Eqs. (\ref{AEQ35}-\ref{AEQ37}) based on the data near critical point.} 
\end{figure*}

\begin{table*}[!t]
\centering
\renewcommand\arraystretch{1}
\small{
\begin{tabular}{lcccl}
\hline
 Interacting system & $\alpha$                      & $\beta$                      & $\gamma$     & $\alpha+2\beta+\gamma$                                                       \\\hline
Kuramoto model & $-1.847\pm 0.035$ ($R^{2}=0.998$) & $1.25\pm0.15$ ($R^{2}=0.924$) & $1.38\pm 0.18$ ($R^{2}=0.942$) & $2.033\pm 0.515$  \\
 \hline
\end{tabular}
}
\caption{\label{Tab3}Critical exponents of the Kuramoto model. We report the estimated value $\pm$ the radius of $95\%$ confidence interval for each critical exponent, accompanied by the fitting accuracy measured by $R^{2}$.}
\end{table*}

The Kuramoto model, a model of coupled phase oscillators, is a representative framework for analyzing synchronization in classical systems \cite{rodrigues2016kuramoto,acebron2005kuramoto}. For convenience, we consider the basic Kuramoto model with $\vert V\vert$ oscillators \cite{rodrigues2016kuramoto,acebron2005kuramoto}
	\begin{align}
		\frac{\partial}{\partial t}\theta_{i}\left(t\right)=\omega_{i}+\frac{K}{\vert V\vert}\sum_{i=1}^{\vert V\vert}\sin\left[\theta_{j}\left(t\right)-\theta_{i}\left(t\right)\right], \label{AEQ47}
	\end{align}
	where $\theta_{i}$ and $\omega_{i}$ denote the phase and natural frequency of the $i$-th oscillator, respectively. Parameter $K$ measures the coupling strength among oscillators. The order parameter of the above interacting system is 
	\begin{align}
		r\left(t\right)\exp\left[\mathsf{i}\psi\left(t\right)\right]=\frac{1}{\vert V\vert}\sum_{i=1}^{\vert V\vert}\exp\left[\mathsf{i}\theta_{j}\left(t\right)\right], \label{AEQ48}
	\end{align}
	in which $\psi\left(t\right)$ is the time-dependent average phase and $r\left(t\right)\in\left[0,1\right]$ measures the synchronization degree of the system. Based on the order parameter, we can reorganize Eq. (\ref{AEQ47}) as
	\begin{align}
		\frac{\partial}{\partial t}\theta_{i}\left(t\right)=\omega_{i}+K r\left(t\right)\sin\left[\psi\left(t\right)-\theta_{i}\left(t\right)\right]. \label{AEQ49}
	\end{align}
	According to Eq. (\ref{AEQ49}), a feedback loop exists between coupling strength $K$ and order parameter $r$, i.e., any increment in $r$ due to the increasing $K$ will enlarge the effective
	coupling among oscillators and attract more oscillators to the synchronous populations in return \cite{rodrigues2016kuramoto,acebron2005kuramoto}.

 To relate the Kuramoto model with our theory, we define a correlation matrix $C$ in terms of synchronization
	\begin{align}
		C_{ab}&=\mathbb{E}\left[r_{ab}\left(t\right)\right], \label{AEQ50}\\
		r_{ab}\left(t\right)\exp\left[\mathsf{i}\psi\left(t\right)\right]&=\frac{1}{2}\left\{\exp\left[\mathsf{i}\theta_{a}\left(t\right)\right]+\exp\left[\mathsf{i}\theta_{b}\left(t\right)\right]\right\}, \label{AEQ51}
	\end{align}
	where the $\left(a,b\right)$-th entity measures the local synchronization degree between oscillators $a$ and $b$. Then we generate the network of oscillators, where there exists an edge between oscillators $a$ and $b$ only if $C_{ab}\simeq 1$, i.e., these two oscillators are strongly synchronous. Here we do not require $C_{ab}= 1$ because numerical errors may lead to a non-one synchronization degree even if oscillators $a$ and $b$ are strictly synchronous (during computation, we determine $C_{ab}\simeq 1$ if $C_{ab}\geq 1-10^{-3}$). Given the network under each condition of $K$, we can measure the size of giant cluster, $M$, and mean cluster size to numerically derive $\Delta S$, $\Delta C$, $\rho_{n}$, and $\chi$.

 \begin{figure*}[t!]
\includegraphics[width=\columnwidth]{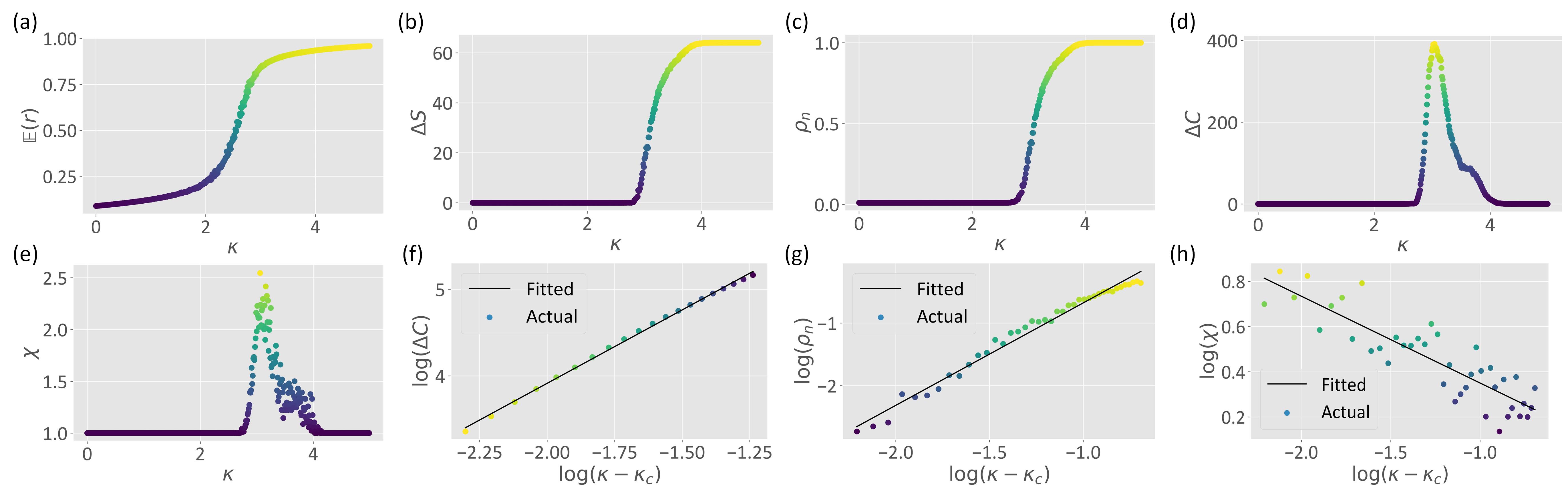}
\caption{\label{G5} Phase transition in the semi-classical Kuramoto model. (a-e) The derived $\mathbb{E}\left(r\right)$, $\Delta S$, $\Delta C$, $\rho_{n}$, and $\chi$ are presented as the functions of $K$. (f-h) A linear fitting is applied to estimate the critical exponents in Eqs. (\ref{AEQ35}-\ref{AEQ37}) based on the data near critical point.} 
\end{figure*}

\begin{table*}[!t]
\centering
\renewcommand\arraystretch{1}
\small{
\begin{tabular}{lcccl}
\hline
 Interacting system & $\alpha$                      & $\beta$                      & $\gamma$     & $\alpha+2\beta+\gamma$                                                       \\\hline
Semi-classical Kuramoto model & $-1.69\pm 0.03$ ($R^{2}=0.998$) & $1.64\pm0.07$ ($R^{2}=0.981$) & $0.38\pm 0.07$ ($R^{2}=0.722$) & $1.97\pm 0.25$  \\
 \hline
\end{tabular}
}
\caption{\label{Tab4}Critical exponents of the semi-classical Kuramoto model. We report the estimated value $\pm$ the radius of $95\%$ confidence interval for each critical exponent, accompanied by the fitting accuracy measured by $R^{2}$.}
\end{table*}

 We define a Kuramoto model with $1000$ oscillators. Each oscillator has a random natural frequency uniformly selected from $\left[1,5\right]$ and exhibits activities for $1000$ time steps. All oscillators are initialized with a random phase selected from $\left[-\pi,\pi\right]$. We set an increasing coupling strength $K\in\left[0,5\right]$ and repeat the experiment under each condition of $K$ for $100$ times. Given each $K$, We denote $\mathbb{E}\left(r\right)$, the mean order parameter derived by time-averaging (i.e, average across time steps) and realization-averaging (i.e., average across repeats), as the expected synchronization degree. In SFigs. \ref{G4}(b-d), the derived $\Delta S$, $\Delta C$, $\rho_{n}$, and $\chi$ are suggested to accurately reflect the disorder-order phase transition around $K\simeq 2.5$ characterized by the sharp increase of $\mathbb{E}\left(r\right)$ in SFig. \ref{G4}(a). In other words, these concepts can reflect the emerging synchronization within the system. Applying the log-log linear fitting on the data near $K\simeq 2.5$, we derive the critical exponents shown in Table. \ref{Tab3} to suggest that the Rushbrooke
inequality in Eq. (\ref{AEQ35}) generally holds.

\subsubsection{Quantum synchronization}\label{ASec2-4}

The elementary Kuramoto model mentioned above only applies to classical systems. To analyze quantum synchronization, we consider the semi-classical Kuramoto model proposed in Ref. \cite{de2014synchronization}. We first reformulate Eq. (\ref{AEQ47}) as a Langevin equation with potential
\begin{align}
    U\left(\theta_{1},\cdots,\theta_{\vert V\vert}\right)=-\sum_{i=1}^{\vert V\vert}\omega_{i}\theta_{i}+\frac{K}{\vert V\vert}\sum_{i,j=1}^{\vert V\vert}\cos\left(\theta_{i}-\theta_{j}\right),\label{AEQ52}
\end{align}
whose quantization defines its semi-classical evolution
\begin{align}
    \frac{\partial}{\partial t}\theta_{i}\left(t\right)=-\frac{U_{i}^{\prime}}{F_{i}}+\frac{\Lambda}{F_{i}}\sum_{j=1}^{\vert V\vert}\left(\beta U_{j}^{\prime}U_{ij}^{\prime\prime}-U_{jji}^{\prime\prime\prime}\right)+\sqrt{\frac{1}{F_{i}}}\xi_{i}, \label{AEQ53}
\end{align}
where we define $F_{i} = e^{-\frac{\Lambda}{D}U_{ii}^{\prime\prime}}$ and $U_{i\cdots k}\equiv\partial_{\theta_i\cdots\theta_k}U$. Notion $\xi$ is a Markovian stochastic fluctuating force
with $\langle\xi_i\left(t\right)\rangle=0$ and $\langle\xi_{i}\left(t\right)\xi_{j}\left(t\right)\rangle= 2\delta_{ij}D\delta\left(t-t^{\prime}\right)$. Notion $\Lambda$ is the quantumness parameter.
The semi-classical Kuramoto model reduces to the original one in Eq. (\ref{AEQ47}) when $\Lambda\to 0$ and $D\to 0$  \cite{de2014synchronization}.

For simulation, we choose $D=\Lambda=1$. The internal frequency and initial phase of each semi-classical oscillator are uniformly distributed in $\left[1,5\right]$ and $\left[-\pi,\pi\right]$, respectively. Due to the high computational complexity of Eq. (\ref{AEQ53}), we set system size as $\vert V\vert=500$. Similar to our experiment in SFig. \ref{G4}, we randomly generate $100$ system realizations to calculate $\mathbb{E}\left(r\right)$, $\Delta S$, $\Delta C$, $\rho_{n}$, and $\chi$ for describing quantum synchronization phase transition (see SFigs. \ref{G5}(a-e) for details). The log-log linear fitting on the data near phase transition derives the critical exponents in Table. \ref{Tab4}, which satisfy the Rushbrooke
inequality in Eq. (\ref{AEQ35}) with acceptable errors.

\subsubsection{Non-linear oscillations with damping}\label{ASec2-5}

Apart from the Kuramoto-type oscillators mentioned above, we also study the Van der Pol oscillator \cite{van1928lxxii}, a mathematical model that describes non-linear oscillations with damping in various electrical circuits and chemical reactions \cite{peltola2007synthesis,ginoux2012van}.  

\begin{figure*}[t!]
\includegraphics[width=\columnwidth]{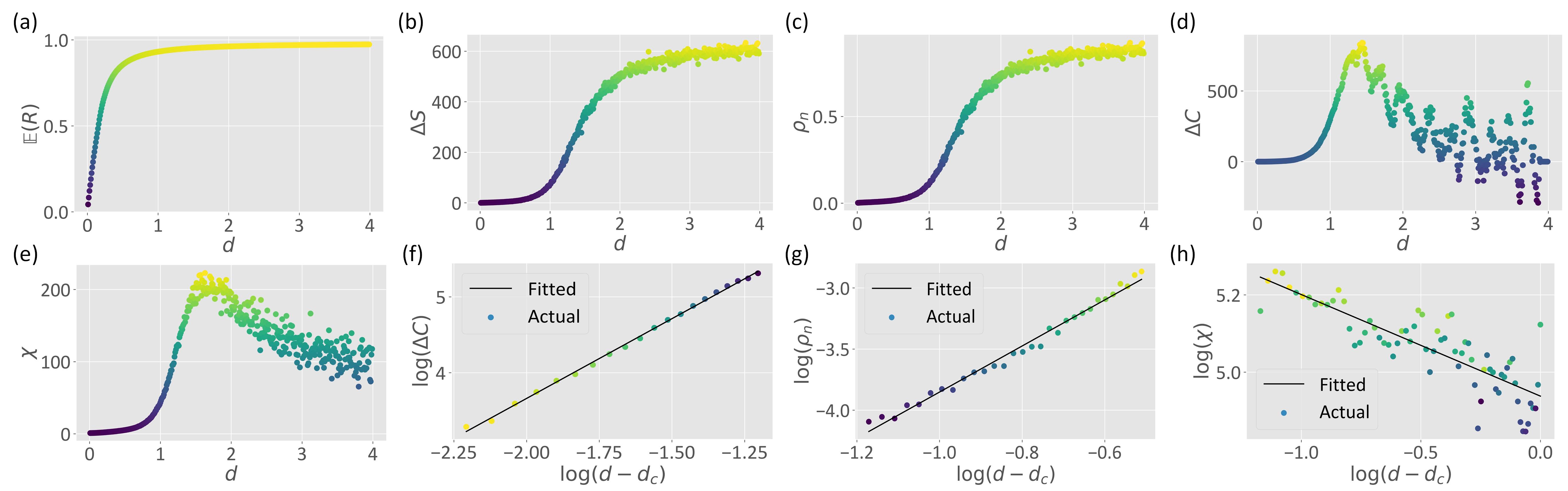}
\caption{\label{G6} Phase transition in the Van der Pol oscillator. (a-e) The derived $\mathbb{E}\left(R\right)$, $\Delta S$, $\Delta C$, $\rho_{n}$, and $\chi$ are presented as the functions of $K$. (f-h) A linear fitting is applied to estimate the critical exponents in Eqs. (\ref{AEQ35}-\ref{AEQ37}) based on the data near critical point.} 
\end{figure*}

\begin{table*}[!t]
\centering
\renewcommand\arraystretch{1}
\small{
\begin{tabular}{lcccl}
\hline
 Interacting system & $\alpha$                      & $\beta$                      & $\gamma$     & $\alpha+2\beta+\gamma$                                                       \\\hline
Van der Pol oscillator & $-2.10\pm 0.04$ ($R^{2}=0.998$) & $1.88\pm0.08$ ($R^{2}=0.987$) & $0.26\pm 0.09$ ($R^{2}=0.713$) & $1.92\pm 0.29$  \\
 \hline
\end{tabular}
}
\caption{\label{Tab5}Critical exponents of the Van der Pol oscillator. We report the estimated value $\pm$ the radius of $95\%$ confidence interval for each critical exponent, accompanied by the fitting accuracy measured by $R^{2}$.}
\end{table*}

Following Refs. \cite{henrici2016synchronization,barron2016stability}, we consider the synchronization of a chain of van der Pol oscillators defined by
\begin{align}
    &\frac{\partial^{2}x_{j}}{\partial t^{2}}+\omega_{j}^{2}x_{j}\notag\\=&2\mu\left(p-x_{j}^{2}\right)\frac{\partial x_{j}}{\partial t}+2\mu d\frac{\partial}{\partial t}\left(x_{j-1}-2x_j+x_{j+1}\right),  \label{AEQ54}
\end{align}
where $1\leq j\leq \vert V\vert$ and parameter $d$ controls the strength of interactions. The periodic boundary condition is considered, i.e., $x_{0}\equiv x_{n}$ and $x_{n+1}\equiv x_{1}$.

In our experiment, we set $p=0.1$ and $\mu=1$. We initialize each $x_{j}$ and $\frac{\partial x_{j}}{\partial t}$ as uniformly distributed in $\left[-1,1\right]$. Meanwhile, we let $\omega_{j}$ be uniformly distributed in $\left[0,1\right]$. We use the average correlation among neighboring oscillators, $\mathbb{E}\left(R\right)$, as the order parameter. To generate the network of oscillators, we initialize a ring graph corresponding to the chain with periodic boundary and remove the edge between each pair of adjacent oscillators if their correlation is smaller than $0.8$. The diluted network is used in subsequent percolation analysis. There are $100$ realizations of the system generated in our experiment, where each realization consists of $1000$ oscillators. The derived results are shown in SFig. \ref{G6} and Table. \ref{Tab5}, validating the Rushbrooke inequality in Eq. (\ref{AEQ35}) with small errors.

 \begin{figure*}[t!]
\includegraphics[width=\columnwidth]{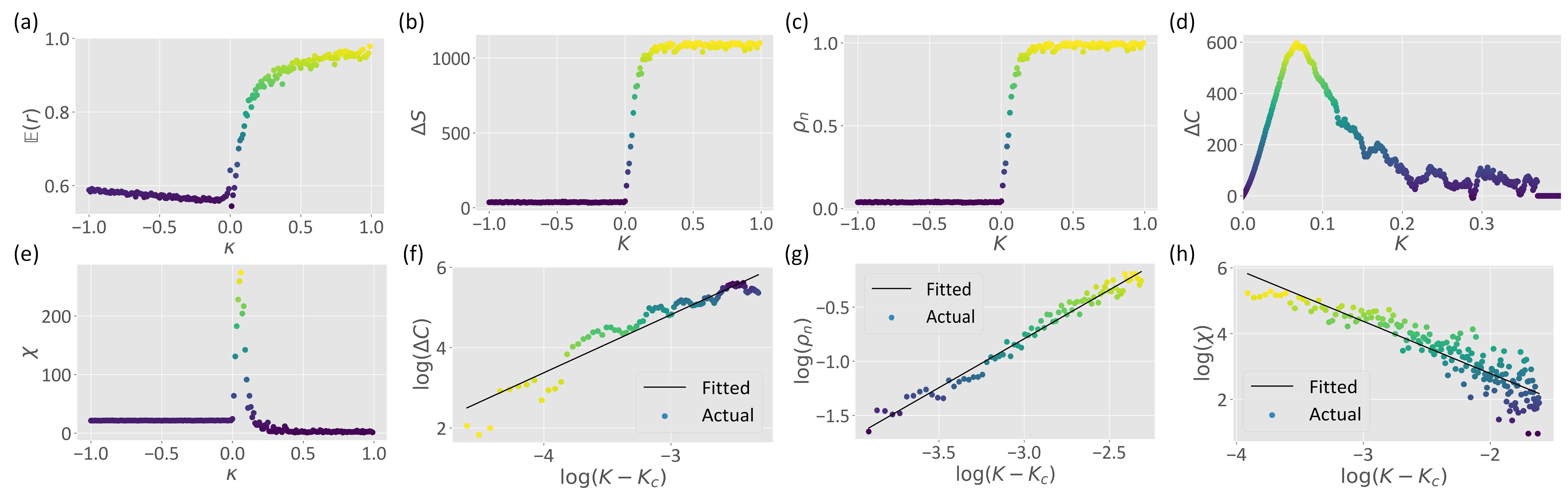}
\caption{\label{G7} Phase transition in the Cellular aggregation and synchronization model. (a-e) The derived $\mathbb{E}\left(r\right)$, $\Delta S$, $\Delta C$, $\rho_{n}$, and $\chi$ are presented as the functions of $K$. (f-h) A linear fitting is applied to estimate the critical exponents in Eqs. (\ref{AEQ35}-\ref{AEQ37}) based on the data near critical point.} 
\end{figure*}

\begin{table*}[!t]
\centering
\renewcommand\arraystretch{1}
\small{
\begin{tabular}{lcccl}
\hline
 Interacting system & $\alpha$                      & $\beta$                      & $\gamma$     & $\alpha+2\beta+\gamma$                                                       \\\hline
Cellular dynamics & $-1.44\pm 0.06$ ($R^{2}=0.947$) & $0.89\pm0.10$ ($R^{2}=0.924$) & $1.58\pm 0.11$ ($R^{2}=0.896$) & $1.92\pm 0.37$  \\
 \hline
\end{tabular}
}
\caption{\label{Tab6}Critical exponents of the cellular dynamics. We report the estimated value $\pm$ the radius of $95\%$ confidence interval for each critical exponent, accompanied by the fitting accuracy measured by $R^{2}$.}
\end{table*}

\subsubsection{Cellular aggregation and synchronization}\label{ASec2-6}
Finally, we consider the cellular aggregation and synchronization modelled by a mixture of the cellular Potts model \cite{unacellular,voss2018cellular} and the swarmalator \cite{sar2022dynamics,hong2023swarmalators}. Please see Ref. \cite{unacellular} for details. In this model, cell interactions depend on clock phases, whose Hamiltonian is formalized as the following
\begin{align}
    H=\sum_{i}\sum_{j\in N_{i}}\left(1-\delta_{\sigma_i,\sigma_j}\right)g\left(\sigma_i,\sigma_j\right)+\lambda\sum_{\sigma_i}\left(a_{i}-A_{i}\right)^2, \label{AEQ55}
\end{align}
where each $i$ denotes a lattice site, notion $N_{i}$ is the set of neighboring sites of $i$, and $\sigma_i$ is the cell at site $i$. Each cell $\sigma_i$ has a volume of $a_{i}$ and a target volume of $A_{i}$. The strength of volume constraint is measured by $\lambda$. Function $g\left(\cdot,\cdot\right)$ denotes the cell-cell adhesion energy between cells
\begin{align}
&g\left(\sigma_{i},\sigma_{j}\right)\notag\\=&\begin{cases}
 J_{0}\left(1-J\cos\left(\theta_{\sigma_{i}}-\theta_{\sigma_{j}}\right)\right),\ &\sigma_{i}\neq 0,\;\sigma_{j}\neq 0 \\
J_{0}, &\sigma_{i}=0\ \text{or} \;\sigma_{j}=0
\end{cases}, \label{AEQ56}
\end{align}
where $\sigma_{i}=0$ stands for that the lattice is not occupied by any cell. Notion $\theta_{\sigma_{i}}$ denotes the internal clock (i.e., the swarmalator \cite{sar2022dynamics,hong2023swarmalators}) defined by 
\begin{align}
    \frac{\partial}{\partial t}\theta_{\sigma_{i}}\left(t\right)=\omega_{i}+\frac{K}{n_{i}}\sum_{j\in N_{i}}\sin\left[\theta_{\sigma_{j}}\left(t\right)-\theta_{\sigma_{i}}\left(t\right)\right],  \label{AEQ57}
\end{align}
where $n_{i}$ is the number of cells located within the neighboring sites of $i$, notion $\omega_{i}$ measures the internal frequency of $\sigma_{i}$, and $K$ denotes the coupling strength.

In our work, we generate $150$ cells on a $40\times40$ square lattice system. We set $\omega_{i}=2$ and $A_{i}=40$ for each cell and define $J=1$ and $J_{0}=1$. Given every control parameter $K\in\left[-1,1\right]$, we repeat $1000$ times of experiments, where we perform $100000$ steps of Metropolis Monte Carlo simulation \cite{geyer2011introduction} and numerically solve Eq. (\ref{AEQ57}) every $10000$ steps in each experiment. In network generation, two neighboring cells are treated as connected if their clock phases are close enough (e.g., the difference between clock phases is smaller than $0.2$ is used in our simulation). The order parameter is defined as $\mathbb{E}\left(r\right)$, the expected synchronization degree derived by time-averaging and realization-averaging. Our derived results are shown in SFig. \ref{G7} and Table. \ref{Tab6}, suggesting that the Rushbrooke inequality in Eq. (\ref{AEQ35}) generally holds.
\end{document}